\documentclass[%
 reprint,
superscriptaddress,
 amsmath,amssymb,
 aps,
 prc,
]{revtex4-2}

\usepackage{amsmath}
\usepackage{textgreek}
\usepackage[utf8x]{inputenc}
\RequirePackage[T1]{fontenc}
\usepackage{graphicx}
\usepackage{dcolumn}
\usepackage{bm}
\usepackage{isotope}
\RequirePackage[sort&compress]{natbib}
\usepackage{subcaption}
\usepackage{mathtools} 
\usepackage{comment}
\RequirePackage[colorlinks,citecolor=blue,urlcolor=blue,linkcolor=blue]{hyperref}

\newcommand{\tj}[6]{ \begin{pmatrix}
  #1 & #2 & #3 \\
  #4 & #5 & #6 
 \end{pmatrix}}

\begin{document}

\preprint{APS/123-QED}


\title{Quantification of spin alignment in fission by simultaneous treatment of gamma and conversion electron angular distributions}

\author{A.~Chalil}
\email{a.chalil@ip2i.in2p3.fr}
\altaffiliation[Present address: ]{Univ. Lyon, Univ. Claude Bernard Lyon 1, CNRS/IN2P3, IP2I Lyon, F-69622, Villeurbanne, France}%
\affiliation{
CEA, DES, IRESNE, DER, Cadarache F-13108 Saint-Paul-Lez-Durance, France
}%
\affiliation{
IRFU, CEA, Universit\'{e} Paris-Saclay, 91191 Gif-sur-Yvette, France
}%

\author{O.~Litaize}%
\affiliation{
CEA, DES, IRESNE, DER, Cadarache F-13108 Saint-Paul-Lez-Durance, France
}%

\author{T.~Materna}
\affiliation{
IRFU, CEA, Universit\'{e} Paris-Saclay, 91191 Gif-sur-Yvette, France
}%

\author{A.~Chebboubi}
\affiliation{
CEA, DES, IRESNE, DER, Cadarache F-13108 Saint-Paul-Lez-Durance, France
}%

\date{\today}

\begin{abstract}
The study of the angular momentum properties of fission fragments can shed light about the complex mechanisms that characterize the fission process. One quantity that is of significant interest, and has not yet been studied adequately, is the alignment of the fragments, which is the cause of anisotropy of the $\gamma$ rays along the fission axis and has been observed in various past and recent experiments. In this work, we have performed calculations using the FIFRELIN code, in an attempt to quantify the alignment of the nuclear spins after neutron-emission. Under the statistical tensor formalism of angular distributions, the conversion-electron and the $\gamma$-ray angular distributions can be treated simultaneously in an event-by-event calculation. This enables a first prediction of the conversion-electron angular distribution with respect to the fission axis. An average value for the alignment of fission fragments is deduced for \isotope[252]{Cf}, with the use of recent experimental data. The method used for the present work can serve as a starting point for future theoretical and experimental studies in terms of $\gamma$ and conversion-electron spectroscopy in view of studying the spin alignment of individual fission fragments, which could further improve our understanding on the process of fission.

\end{abstract}

\maketitle


\section{Introduction}
\label{intro}

From its discovery during the late 30's~\cite{Hahn1939,Meitner1939} up to today, nuclear fission has been continuously and extensively studied, both for its significance in Nuclear Physics and Astrophysics~\cite{Andreyev2018-ks,Schmidt_2018}, as well as its role in various applications such as in energy production.  Despite the continued efforts, not all aspects of the particular phenomenon are fully understood. Experimental observables from the fission fragment de-excitation, either by neutron, $\gamma$ or conversion-electron emission can shed light on the underlying mechanisms of the fission process and lead to a better understanding of the phenomenon. Besides the need to pursuit a complete theoretical understanding of the fission process, exact predictions of the $\gamma$ ray yields are essential for technological applications in nuclear reactors~\cite{RIMPAULT20123, Lemaire_2015}.

Experimentally, a plethora of observables can contribute to the understanding of fission. Average numbers and energies of the emitted $\gamma$ rays have been deduced in~\cite{Pleasonton_PhysRevC.6.1023} for the thermal-neutron-induced fission  of \isotope[235]{U}.  Precise product mass yields for \isotope[235]{U} have been recently measured~\cite{Chebboubi2021} using the LOHENGRIN spectometer~\cite{ARMBRUSTER1976213} at ILL, Grenoble. Spontaneous fission (SF) of \isotope[252]{Cf} has first been studied in~\cite{Verbinski_PhysRevC.7.1173}, measuring the average and total $\gamma$ ray energy released per fission. Distributions of energies and multiplicities have also been measured in~\cite{Chyzh_PhysRevC.85.021601}
using the DANCE array~\cite{HEIL2001229}, indicating the stochastic nature of the emitted $\gamma$ rays. These observables where measured again in~\cite{Oberstedt_2015_PhysRevC.92.014618} with higher precision using an artificial diamond detector~\cite{OBERSTEDT201331}.

A cumulative interest in the recent years has been focusing on angular momentum studies of the fission process. The first study that focused on observables related to angular momenta has been carried out in~\cite{Wilhelmy_PhysRevC.5.2041}, using measurements of angular correlations between the fission fragments and the subsequent $\gamma$ rays. Measurements of the intensities of the low-lying ground-state band transitions $2^+ \rightarrow 0^+$ were forward-peaked with respect to the fission axis, which is evidence that angular momentum is aligned perpendicularly to the direction of the fragment (perpendicular alignment). This is also supported in various different studies~\cite{Skargvag_1980_PhysRevC.22.638,Wolf_1976_PhysRevC.13.1952,NIX19651,Hoffman_PhysRev.133.B714}. The initial alignment is gradually destroyed by the de-excitation of the fragment by n/ $\gamma$ / e$^-$ emission. After neutron emission, the distribution of $m$-substates can be approximated by a gaussian distribution centered on zero as in~\cite{Hoffman_PhysRev.133.B714}. 

\begin{figure*}[t]
    \centering
    \includegraphics[width=\textwidth]{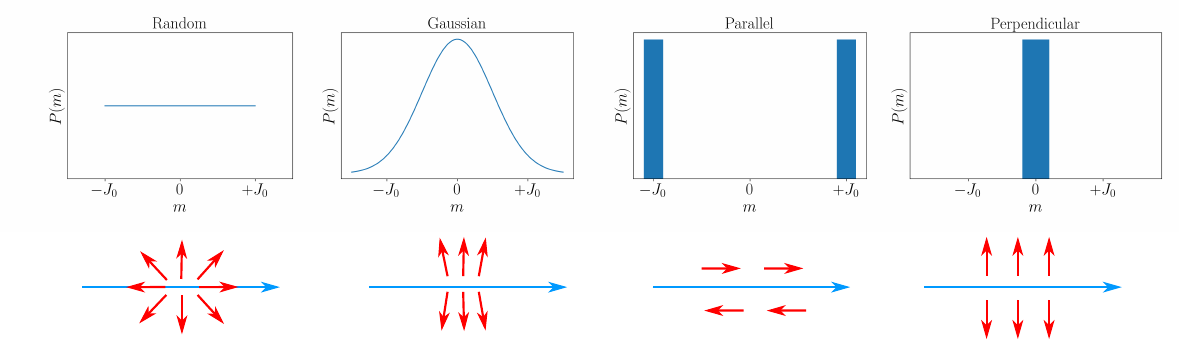}
    \caption{Schematic illustration for various ways of alignment of the spins along a quantization axis. See text for details.}
    \label{fig: alignment}
\end{figure*}
Furthermore, in~\cite{Smith_PhysRevC.60.064611}, measurements
of angular directional correlations between
the $\gamma$ rays emitted from one fragment with the $\gamma$ from the complementary fragment, for decays from low-lying excited states were presented for the first time for the SF of \isotope[252]{Cf}. A recent study~\cite{Wilson2021} confirms no dependence between the spin and the mass or charge of the fission fragments. In the same study, no correlation between the spin values between the light and the heavy fragment was found, suggesting that angular momentum is generated via two independent torques during scission. These conclusions are discussed in~\cite{Randrup_2021_PhysRevLett.127.062502}, where it is shown using the nucleon-exchange mechanism that uncorrelated spins are not necessarily generated after scission.

In~\cite{Bertsch_2019_PhysRevC.99.034603}, a detailed study of the effects of the nuclear deformation on the angular momentum distribution of the fission fragments has been presented, by modeling the angular distribution of $\gamma$ rays the fission fragments from a fully vertically aligned initial spin population of the fragments. Branching ratios in induced fission are investigated in~\cite{Bertsch_2020_PhysRevC.101.034617}, using a schematic model and in~\cite{Bertsch_2023_PhysRevC.107.044615} by using a configuration-interaction model.

Furthermore, a new method of probing correlated fission fragment spins from helicity measurements is discussed in~\cite{Randrup_2022_PhysRevC.106.014609}. The first unrestricted microscopic calculations of the primary fission fragments' intristics spins and their orbital angular momentum have also been realised in~\cite{Bulgac_2022_PhysRevLett.128.022501}.

In this work, an estimation on the degree of the spin alignment of the fission fragments on the fission axis after neutron emission is presented, for the case of SF of \isotope[252]{Cf}. For this estimation, a Monte Carlo approach with the aide of the FIFRELIN code~\cite{Litaize2015} has been used. In order to constrain the alignment, experimental data on the angular distribution of $\gamma$ rays \cite{oberstedt_2018} have been used. We demonstrate the ability of the FIFRELIN code to tune the initial alignment in order to describe the experimental data, which can be coupled with other variables, such as ratios of the multipolarities of the transitions, such as the E2/M1 ratio studied in~\cite{oberstedt_2018}.

The spin alignment after neutron emission can have various forms, depending on the orientation of the spin vectors along the quantization axis~\cite{STUCHBERY200369,Heather_2003_PhysRevC.68.044312}. For the case of fission, this quantization axis corresponds to the fission axis. In Fig.~\ref{fig: alignment}, a schematic illustration of some cases of spin orientations are shown. For the case of a randomly oriented spin distribution, all magnetic substates are equiprobable. For a gaussian distribution, the magnetic substates tend to be oriented perpendicularly on the fission axis depending on the standard deviation of the gaussian. For the case of a fully parallel alignment, all spins of the fragments are aligned parallel to the fission axis, meaning that only the magnetic substates with the maximum absolute value are populated. The opposite is a fully perpendicular alignment where all the spins are perpendicular to the fission axis and only the magnetic substates with the minimum absolute value of $m$ are populated. The vertical orientation of spins correspond to the wriggling/bending modes, while the parallel orientation to the tilting/twisting modes, which are described in~\cite{Randrup_2022_PhysRevC.106.014609}.

For the case of fission and after neutron emission, a partially perpendicular alignment can be approximated with the gaussian distribution centered at $m=0$. The gaussian distribution has also been used to describe the alignment of product nuclei in fusion-evaporation reaction after neutron-emission~\cite{Diamond_1966_PhysRevLett.16.1205,YAMAZAKI19671}.

The statistical tensor formalism~\cite{Rose_Brink_Revmodphys_1967,Hamilton1975_electromagnetic}, which has been implemented for the description of the $\gamma\gamma$ angular correlations in FIFRELIN~\cite{Chalil2022} has been used for the present study. For the present calculations, the initial distribution of the $m$-substates of each fragment was assumed to follow a gaussian distribution centered at $m=0$. The standard deviation of the gaussian distribution was varied in order to match the experimental data from~\cite{oberstedt_2018}. In addition, it was possible to implement the directional correlations for conversion electrons, in an attempt to obtain the first prediction of the fission fragment-conversion-electron angular distribution. Both distributions are treated simultaneously in a single calculation. 

The deduced degree of alignment is directly related to the population of the magnetic substates of the fragments along the fission axis and can provide with useful input on theoretical models and constrain their parameters. The details of the method are described in the following section. 

\section{Methods}

The code FIFRELIN, developed at CEA Cadarache in order to accurately model the fission process is used for the present study. In a recent work, FIFRELIN was capable of reproducing the experimental neutron- and gamma multiplicities by using an energy dependent spin cut-off model to account for the initial total angular momentum distribution, in combination with microscopic level density models~\cite{PIAU2023137648}. In addition, the ability to simulate the effect of $\gamma$-directional correlations has been also recently implemented in the FIFRELIN code~\cite{Chalil2022}. The method used the statistical tensor formalism~\cite{Rose_Brink_Revmodphys_1967,Hamilton1975_electromagnetic, STUCHBERY2002753,ROBINSON2002469}, which are quantities directly related to the populations of the magnetic substates of the initial spin state. This method opens the way for studies using different kinds of spin alignment on the fission axis for each fragment. An unequal population of these states along a quantization axis results in an anisotropic emition of radiation. In~\cite{Chalil2022}, the method was applied by assuming that the initial state is unoriented, which corresponds in an initial condition for the statistical tensor:
\begin{equation}\label{eq: unoriented statistical tensor}
    \rho^{\lambda}_q (J_0)= \delta_{\lambda 0} \delta_{q 0},
\end{equation}
where $\lambda$ is the rank of the statistical tensor and $q$ is an integer within $-\lambda \leq q \leq \lambda$ and $\delta$ is the Kronecker delta.  In most cases however, the initial state is not unoriented. Examples of such initial unoriented states can be formed after a reaction, where the ejectile has enough energy to form higher waves than the $s$ wave, or after the detection of a previously emitted particle~\cite{Hamilton1975_electromagnetic}. The detection of the fission fragment will generally result in an anisotropic emission of the prompt $\gamma$ rays emitted in the fission process. The statistical tensor is directly related to the population of the $m$ substates along an axis of quantization, which in our case is the fragment axis, by the following relation~\cite{Hamilton1975_electromagnetic}:
\begin{equation}\label{eq: oriented statistical tensor}
    \rho^{\lambda}_0 (J_0)= \sum_{m} (-1)^{m+J_0} (2\lambda +1)^{1/2} \tj{J_0}{J_0}{\lambda}{-m}{m}{0} P(m) ,
\end{equation}
where $J_0$ is the spin of the initial state, $m$ are the magnetic substates, $\tj{J_0}{J_0}{\lambda}{-m}{m}{0}$ is a Wigner-$3j$ symbol and $P(m)$ is the probability distribution function of the populations of these substates. 


In this work, we consider as initial state $J_0$ the first state that the fragment has after neutron emission and a gaussian distribution, centered at zero, of the form:
\begin{equation}\label{eq: gaussian distro}
   P(m) = \frac{1}{2 \sigma \sqrt{2 \pi}} \exp (- \frac{m^2} {2\sigma^2}  ) 
\end{equation}
where we choose the simple form for $\sigma=a  J_0$. This form has also been used in~\cite{Hoffman_PhysRev.133.B714}, in order to determine the degree of alignment for fission fragments after the fission of \isotope[233,235]{}{U} and \isotope[239]{}{Pu}, using analytical methods. The determination of the factor $a$ is directly connected with the alignment averaged in all fission fragments, as the fission fragments have various spins described from spin distribution functions.

During fission and after neutron emission, the fragments emit $\gamma$ rays but also conversion electrons. The directional correlation of conversion electrons during the de-excitation process was not considered in~\cite{Chalil2022}. Although most of the conversion electrons are usually emitted at the end of the cascades, which could impact weakly the distribution of the $\gamma$ rays, it is necessary to implement the directional correlations of the conversion electrons using the statistical tensor formalism, as is done with the $\gamma$ rays. This will enable a first prediction of the conversion $e^-$ angular distribution which can be a significant observable for angular momentum studies. In order to achieve this, when a conversion electron is emitted, the angular distribution coefficient~\cite{Hamilton1975_electromagnetic} has to be modified as:
\begin{align}\label{eq: generalized ang_distro coeff}
    A^{\lambda_i \lambda_f}_\lambda(J_f,J_i,L,L') =& \frac{1}{1+\delta^2} 
     [ b_{\lambda}(L,L) F^{\lambda_i \lambda_f}_\lambda(L,L,J_f, J_i) \nonumber \\ &+ 2\delta b_{\lambda}(L,L') F^{\lambda_i \lambda_f}_\lambda(L,L'J_f, J_i) \nonumber \\ &+  \delta^2 b_{\lambda}(L',L') F^{\lambda_i \lambda_f}_\lambda(L',L',J_f, J_i) ],
\end{align}
where $F$ is the generalized angular distribution coefficient~\cite{Hamilton1975_electromagnetic,Ferentz_F_coeff}. The difference between the treatment of $\gamma$ rays and conversion electrons within the angular distribution coefficient is the inclusion of the factor $b_\lambda$. This factor is called \textit{particle parameter} and their calculation can be quite complicated, as they depend also on nuclear structure effects~\cite{HAGER19691_part3}. Fortunately, for conversion electrons, these parameters have been calculated to an extensive range of elements in~\cite{HAGER1968397}, which were the values used for this work in order to implement the directional correlation of the conversion electrons. The range of these calculations, within $Z=30-103$ includes the fragments emitted during fission of \isotope[252]{}{Cf}. 

\begin{figure}[t]
    \centering
\includegraphics[width=\columnwidth]{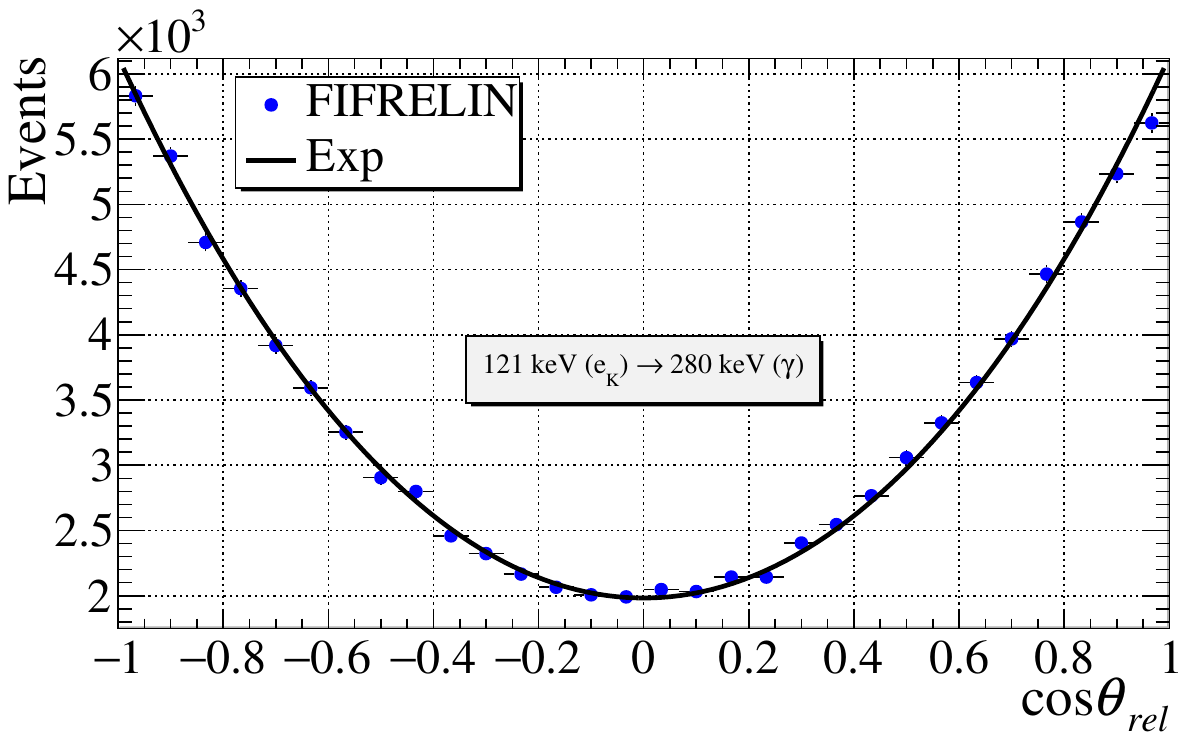}
    \caption{Simulated conversion $e^-$-$\gamma$ angular correlation from the decay of \isotope[75]{As}. The curve represents the distribution from the experimentally determined angular distribution coefficients in~\cite{RAESIDE1969677}}
    \label{fig: 75Ar egamma corr}
\end{figure}

In order to demonstrate the correct implementation, a conversion electron-$\gamma$ ray angular correlation from the decay of \isotope[75]{As} is compared with the experimental result in Fig.~\ref{fig: 75Ar egamma corr}.
\begin{figure*}[ht]
	\centering
	\begin{subfigure}{0.5\textwidth}
		\includegraphics[width=\textwidth]{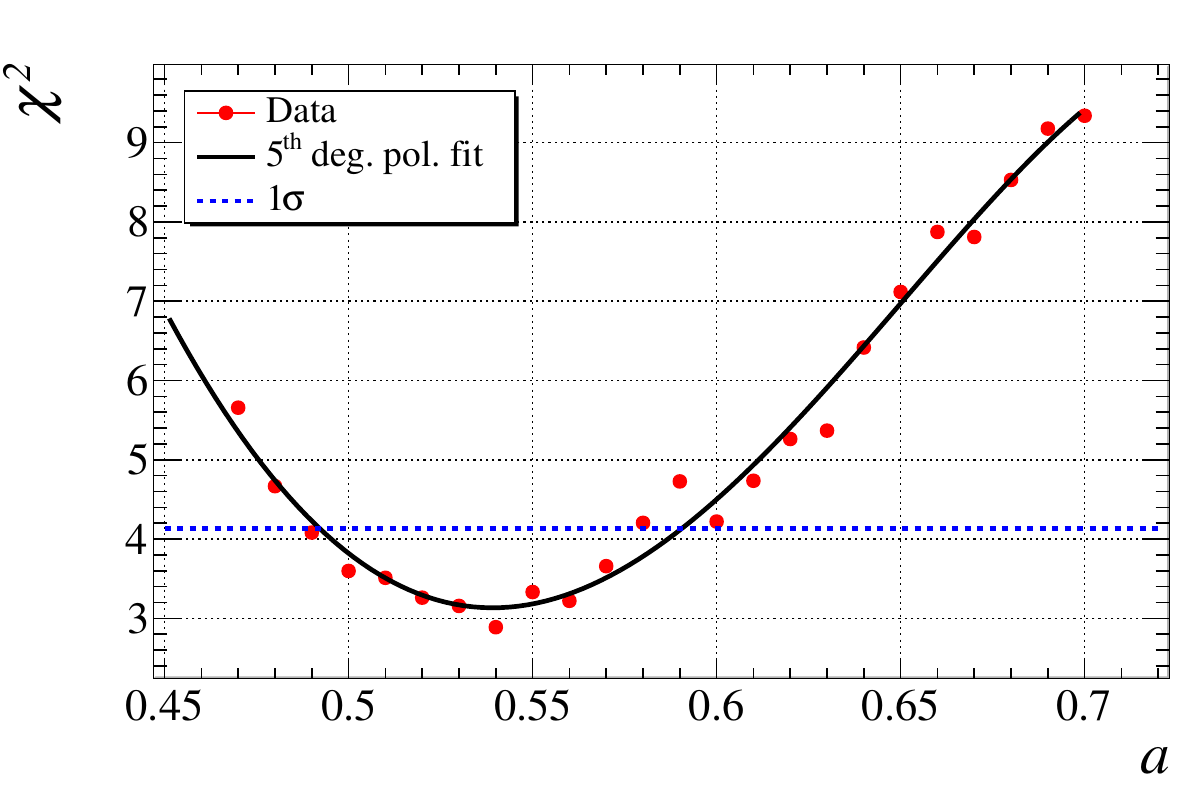}
		\centering
		\caption{}
        \label{subfig: minimization}
	\end{subfigure}%
	\begin{subfigure}{0.5\textwidth}
		\includegraphics[width=\textwidth]{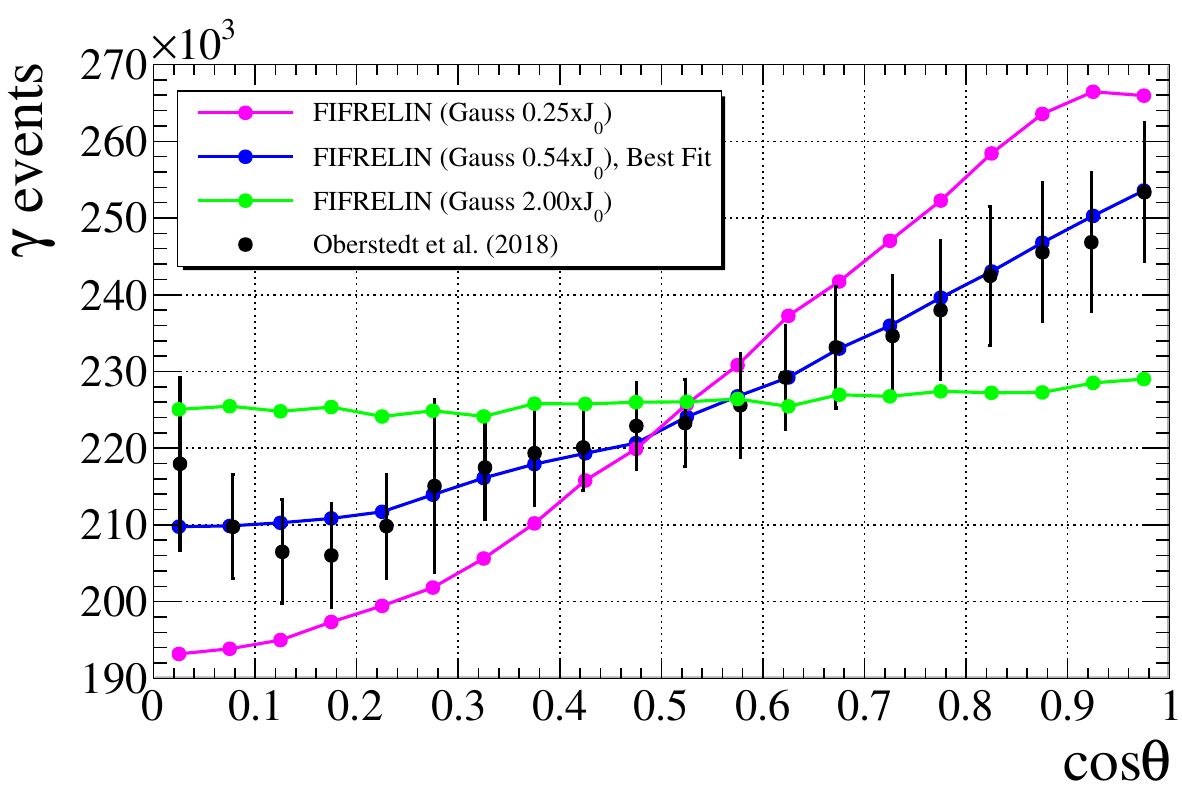}
		\centering
		\caption{}
      \label{subfig: best fifrelin}
	\end{subfigure}%
 \caption{(a) Minimization of the parameter $a$ between data from~\cite{oberstedt_2018} and FIFRELIN calculation. (b) FIFRELIN calculation of the fission fragment-$\gamma$ angular correlation using the optimal value of $a$ (blue curve). Two more cases for $a=0.25$ (purple curve) and $a=2$ (green curve) are also shown. Experimental data are shown in black. See text for details.}
   \label{fig: minization}
\end{figure*}
The experimental curve is corrected for solid angle effects in~\cite{RAESIDE1969677}. The result for the 121 ($e^-$) $\rightarrow$ 280 ($\gamma$) keV cascade  show that FIFRELIN is able to simulate precisely the correlations involving conversion electrons.  

The next step is to determine the emission angles of the $\gamma$ rays and conversion electrons using the same method as in~\cite{Chalil2022} but with the initial condition of Eq.~\ref{eq: oriented statistical tensor} and the modified angular distribution coefficient of Eq.~\ref{eq: generalized ang_distro coeff} when a conversion electron is emitted during the de-excitation of the fission fragments of \isotope[252]{}{Cf}.

\section{Results and Discussion}

A million spontaneous fission processes of \isotope[252]{Cf} were generated using the FIFRELIN code. Typically a fragment will emit 2-3 neutrons and after will de-excite by $\gamma$ or conversion $e^-$ emission. For this work, the direction of neutrons along the fission axis was considered isotropic. After the emission of the last neutron, the initial condition of Eq.~\ref{eq: oriented statistical tensor} was set on the first state of the heavy and light fragments.

During the event generation, the emission angles of the $\gamma$ rays and conversion electrons were treated simultaneously, allowing us to obtain both distributions within a single calculation. In order to tune the alignment, a $\chi^2$ minimization was necessary in order to find the value of $a$ that describes best the data published in~\cite{oberstedt_2018}. The $\chi^2$ function was formed:
\begin{equation}\label{eq:S_squared}
	\chi^2 = \sum_i \left[ \frac{W_{i}(\theta^i)-W^i_{FIF}(a,\theta^i) }{\sigma_{W_{exp}}  } \right]^2,
\end{equation}
where $W_{i}(\theta^i)$ are the experimental data, $\sigma_{W_{exp}}$ is the error of the data and $W^i_{FIF}(a,\theta^i)$ are the theoretical values generated with FIFRELIN. The parameter $a$ in Eq.~\ref{eq: gaussian distro} was varied in a small step until a minimum was reached.
The results of the minimization are shown in Fig.~\ref{fig: minization}. In Fig.~\ref{subfig: minimization}, the value of Eq.~\ref{eq:S_squared} is shown as a function of the parameter $a$. The optimal value of $a$ is found equal to $0.54(5)$. The error on this value was assigned using the relation~\cite{ROBINSON1990386}:
\begin{equation}\label{eq:internal_error}
\chi_{lim}^2 = \chi^2_{min} +1,
\end{equation}
It is important to note that the assigned uncertainty is only the statistical uncertainty and does not include any uncertainties arising from the unknown multipolarities and spins of the sampled states. The minimization was performed in the whole angle range as systematic errors are included for small and very large angles as the authors discuss in~\cite{oberstedt_2018}.

Two more cases FIFRELIN calculations are shown in Fig.~\ref{subfig: best fifrelin}. For $a=0.25$, the shape of the gaussian is more sharp, leading to significant population of the low  values $m$-distribution close to $m=0$. This results in higher vertical alignment, leading to a higher slope of the angular distribution. Contrariwise, for $a=2$, the sigma of the gaussian is larger, leading to more $m$ substates populated significantly. The slope of the distribution is signifantly decreased, approximating a uniform distribution. This agrees with the principle that equal population of $m$-substates will result in isotropic emission of the radiation~\cite{Rose_Brink_Revmodphys_1967,Hamilton1975_electromagnetic}. This effect is nicely demonstrated with the present results.

\begin{figure}[t]
    \centering
\includegraphics[width=\columnwidth]{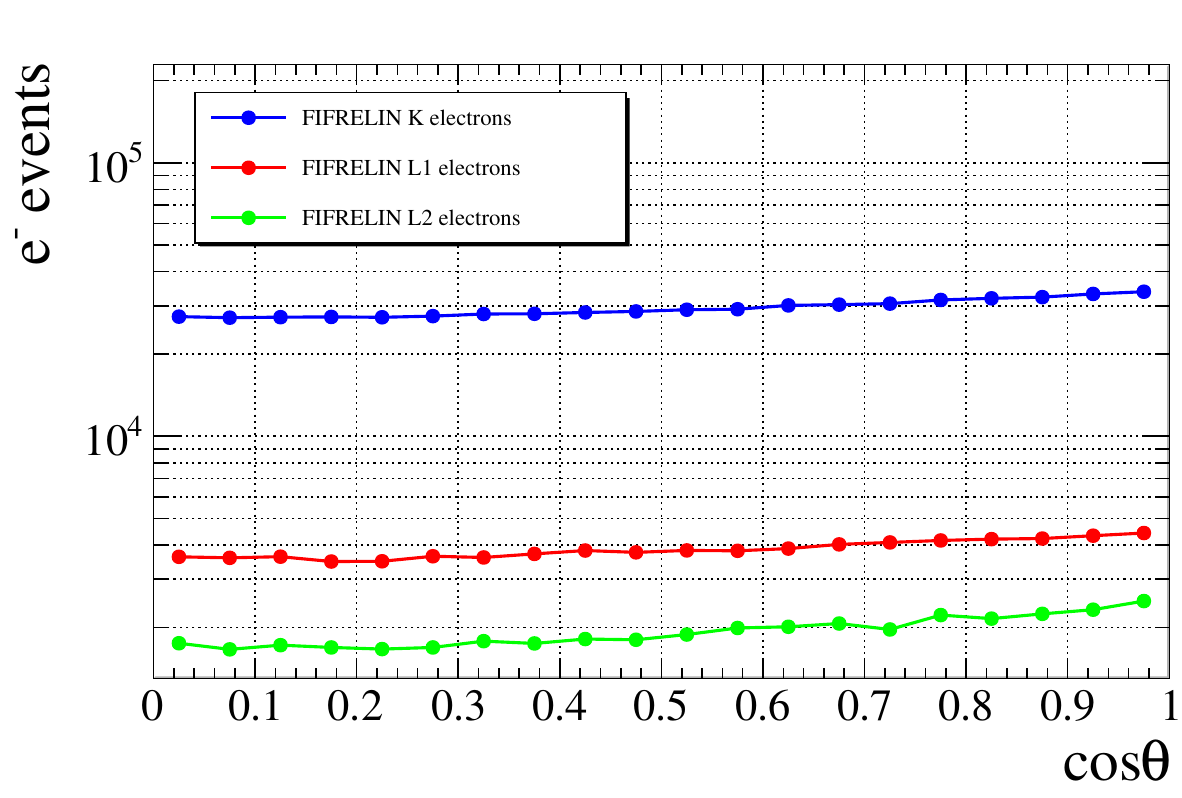}
    \caption{The predicted angular distribution of $K, L1$ and $L2$ conversion electrons with respect to the fission axis, derived for $a=0.54$, which corresponds the best description of angular distribution of the $\gamma$ rays, using the FIFRELIN code.}
    \label{fig: e distro}
\end{figure}

\begin{figure}[t!]
    \centering
    \includegraphics[width=\columnwidth]{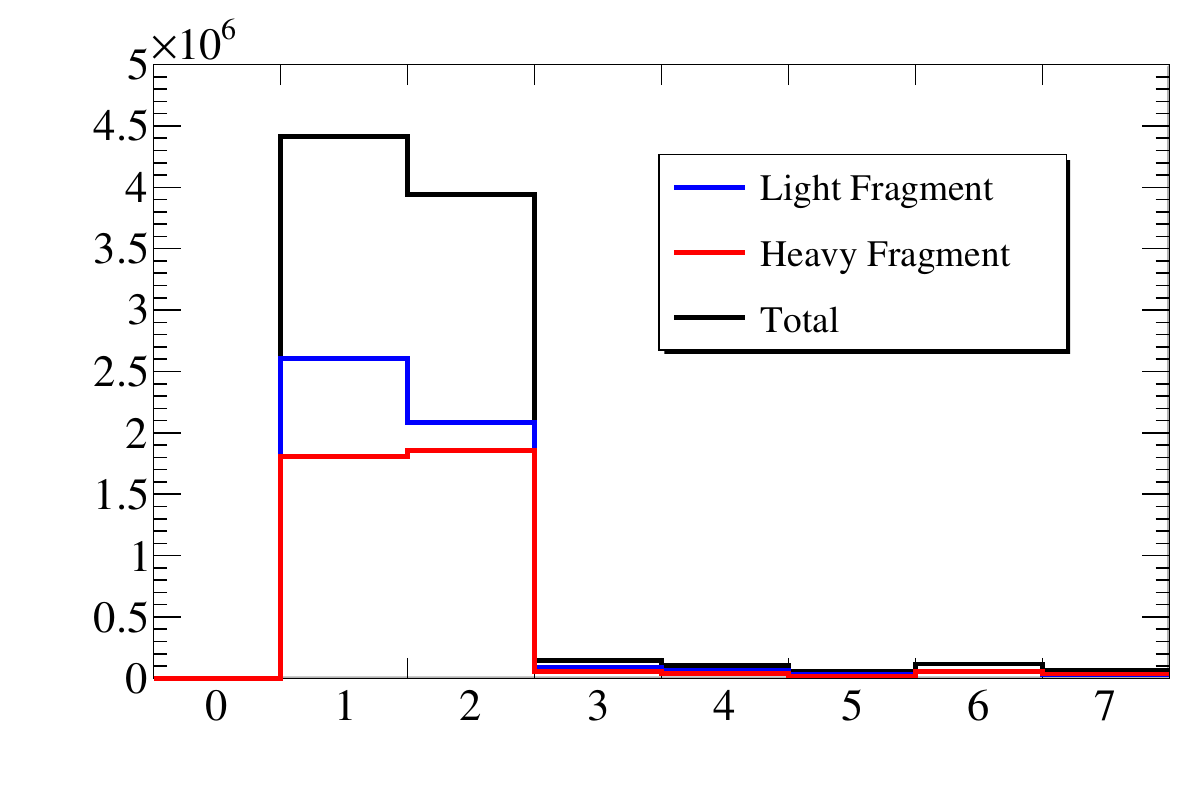}
    \caption{Order of multipoles emitted during 1 million fission events by the FIFRELIN code. The light and heavy fragment components are also shown.}
    \label{fig: multipolarities}
\end{figure}

The fit of the alignment on the $\gamma$ ray experimental data enables the present calculation to simultaneously predict the total conversion $e^-$ angular distribution. In Fig.~\ref{fig: e distro}, the angular distribution of $K, L1$ and $L2$ conversion electrons emitted after 1 million fissions of \isotope[252]{Cf} is shown, as calculated using the statistical tensor formalism within the FIFRELIN code. These electrons account for almost $90\%$ of the emitted conversion electrons in FIFRELIN. It is to be noted that the addition of the $L1$ and $L2$ result in a $2\%$ difference on the value of the parameter $a$. The electrons emitted from other atomic shells are considered negligible in the present analysis.

In order to compare the slopes of each of the distributions, the angular correlation function:
\begin{equation}\label{eq: ang cor function}
    W(\cos \theta) = A_0 \left( a_2 P_2 (\cos \theta) + a_4 P_4 (\cos \theta) \right)
\end{equation}
can be fitted to the FIFRELIN calculations for the $\gamma$ and the $K, L1$ and $L2$ conversion electrons. The coefficients $a_2$ and $a_4$ are tabulated in Table~\ref{tab: ang cor coeffs}.

\begin{table}[t]
    \centering
    \begin{tabular}{c|c|c }
    
        Particle & $a_2$ & $a_4$ \\
        \hline
    \hline
         $\gamma$ & 0.126(2) & -0.005(4) \\
         $e^-_K$ & 0.158(3) & 0.002(4) \\ 
          $e^-_{L1}$ & 0.160(8) & -0.006(11) \\
           $e^-_{L2}$ & 0.272(12) & 0.015(16)
    \end{tabular}
    \caption{Coefficients $a_2$ and $a_4$ obtained from fitting the angular correlation function of Eq.~\ref{eq: ang cor function} to the FIFRELIN calculations for $\gamma$ rays and conversion electrons. Errors are statistical, corresponding to 1 million fission events.}
    \label{tab: ang cor coeffs}
\end{table}

It is important to note that this new method of determining the alignment depends heavily on multipolarities of transitions during the de-excitation of the fragments. In the present work, the multipolarities of RIPL-3 have been used within FIFRELIN, as well as the multipolarity mixing ratios for ENSDF. In Fig.~\ref{fig: multipolarities}, a histogram displaying the order of multipoles for 1 million fission events is shown, for both the light and the heavy fragments. The ratio of the Quadrupole (L=2) to Dipole (L=1) transition is found to be 0.89. 
Since most of these transitions are of electric nature (E1,E2), this could explain the larger $a_2$ values for the conversion-electron distributions compared to the value of the $\gamma$ ray distributions. The particle parameter $b_2$ is larger than one for these type of transitions, leading to a higher slope of the electron distribution compared to the $\gamma$ ray one. The higher $a_2$ for $L2$ electrons can be explained by the higher $E2/M1$ ratio of the emitted electrons for each shell, where it is found that $\delta(E2/M1;L2) = 2.00$ compared to $\delta(E2/M1;L1)=0.97$ and $\delta(E2/M1;K)=1.15$. It is important to note however, that since the present calculations predict the total angular distribution of electrons and gammas, other factors can significantly influence the slope of the distributions, such as the different combinations of multipolarities and spins together with the number of the electron shell from where the electron is emitted.

A source of uncertainty comes also from the transitions starting from high excitation energies, of which the multipolarities have not been determined experimentally. In the present work, such transitions were treated as pure transitions with zero mixing. Although transitions from the continuum usually decay as pure dipole transitions, small mixings may affect the final results. Furthermore, isomeric states may also reduce the slope of the angular distributions, due to the attenuation of the angular correlation coefficients. This could result in a lower value of the parameter $a$. Furthermore, in the present work, all $\gamma$ rays and conversion $e^-$ which are predicted are treated in order to extract the total angular distribution. Experimentally the detection of a proportion of these radiations is not possible, due to the detection limits of the various setups.

The result is differs from previous values reported in~\cite{Hoffman_PhysRev.133.B714} for the fission of different nuclei: \isotope[233]{U}, \isotope[235]{}{U} and \isotope[239]{Po}, where the value of $a$ lies close to 0.3. The authors however state that their results apply only for these isotopes with neutron multiplicities 1.23,
1.22, and 1.45, respectively. They have also considered only dipole and quadrupole transitions and there is no treatment of conversion electrons. The difference could be attributed to the higher neutron multiplicity of \isotope[252]{Cf}, which is calculated with FIFRELIN to 2.066 for the light and 1.61 for the heavy fragment. Higher neutron multiplicity results in further de-orientation of the spins, which decreases the slope of the angular distribution.

There is the possibility to constrain both the multipolarities, such as the total quadrupole to dipole ratio and the alignment using the present method. This requires experimental data for the angular distribution of conversion electrons with respect to the fission axis, which are currently missing in literature. Using the experimental data on both electron and $\gamma$ ray distributions would enable the minimization of both the $a$ parameter and the total E2/M1 mixing with stricter constraints.

\section{Conclusions and future directions}

Overall, the present work presents a novel approach on determining the mean alignment of the fission fragments after neutron emission by accounting both $\gamma$ and conversion-electron emission, using the Monte Carlo code FIFRELIN. The functionality implemented in FIFRELIN in order to calculate angular distributions depending on the initial alignment of spins along the fission axis and the multipolarities involved can be useful for angular momentum studies in fission. A possibility to implement the tensor formalism for neutron angular distribution is a direction that would allow also a more complete picture of the destruction of the initial alignment during neutron emission, which could be significant due to the large angular momentum removal~\cite{Stetcu_PhysRevLett.127.222502}. 

The exact knowledge of multipolarities of the radiations involved during fragment de-excitation is also of great importance, as it can affect the shape of the angular distribution. Experimental values of multipolarity mixing ratios are largely missing from databases, and experimental studies for their determination are necessary input for constraining theoretical models. 

The knowledge of the distribution of the $m$-substates can also set constraints on the amount of the wriggling/bending and tilting/twisting modes of angular momentum generation in fission. For a complete picture though, neutron-emission has to be treated also within the present formalism, in order to treat the cascades immediately after fission. The statistical tensor formalism can also be generalized to include polarization~\cite{Hamilton1975_electromagnetic}, which can separate the components of parallel and anti-parallel spins in a single calculation.

The angular distribution of conversion electrons with respect to the fission axis is also a very important observable that has not been studied thoroughly, possibly because of the complicated determination of the particle parameters used in Eq.~\ref{eq: generalized ang_distro coeff}. The tabulated values from~\cite{HAGER1968397} have been used in this work in order to simulate the electron distributions. This allowed a simultaneous calculation of both electron and the $\gamma$ distributions depending on the initial alignment.

A measurement of the fragment-electron angular distribution seems essential input for future studies, in order to minimize the uncertainties arising from the unknown multipolarities of the transitions, especially those that depopulate higher energy states of the fragments. Potential experimental campaigns are necessary, and will give essential information and input, in order to constrain both alignment and multipolarities, and allow for a more complete description of the fission fragment de-excitation, while at the same time give useful information on the evolution of the angular momentum properties.

\section*{Acknowledgments}

We acknowledge the financial support of the Cross-Disciplinary Program on Numerical Simulation of CEA, the French Alternative Energies and Atomic Energy Commission. A. Chalil is grateful to Prof. Theo J. Mertzimekis for the useful discussions and comments on the present work.

\bibliographystyle{apsrev4-1}
\bibliography{fission.bib,stereo.bib}

\begin{thebibliography}{45}%
\makeatletter
\providecommand \@ifxundefined [1]{%
 \@ifx{#1\undefined}
}%
\providecommand \@ifnum [1]{%
 \ifnum #1\expandafter \@firstoftwo
 \else \expandafter \@secondoftwo
 \fi
}%
\providecommand \@ifx [1]{%
 \ifx #1\expandafter \@firstoftwo
 \else \expandafter \@secondoftwo
 \fi
}%
\providecommand \natexlab [1]{#1}%
\providecommand \enquote  [1]{``#1''}%
\providecommand \bibnamefont  [1]{#1}%
\providecommand \bibfnamefont [1]{#1}%
\providecommand \citenamefont [1]{#1}%
\providecommand \href@noop [0]{\@secondoftwo}%
\providecommand \href [0]{\begingroup \@sanitize@url \@href}%
\providecommand \@href[1]{\@@startlink{#1}\@@href}%
\providecommand \@@href[1]{\endgroup#1\@@endlink}%
\providecommand \@sanitize@url [0]{\catcode `\\12\catcode `\$12\catcode
  `\&12\catcode `\#12\catcode `\^12\catcode `\_12\catcode `\%12\relax}%
\providecommand \@@startlink[1]{}%
\providecommand \@@endlink[0]{}%
\providecommand \url  [0]{\begingroup\@sanitize@url \@url }%
\providecommand \@url [1]{\endgroup\@href {#1}{\urlprefix }}%
\providecommand \urlprefix  [0]{URL }%
\providecommand \Eprint [0]{\href }%
\providecommand \doibase [0]{http://dx.doi.org/}%
\providecommand \selectlanguage [0]{\@gobble}%
\providecommand \bibinfo  [0]{\@secondoftwo}%
\providecommand \bibfield  [0]{\@secondoftwo}%
\providecommand \translation [1]{[#1]}%
\providecommand \BibitemOpen [0]{}%
\providecommand \bibitemStop [0]{}%
\providecommand \bibitemNoStop [0]{.\EOS\space}%
\providecommand \EOS [0]{\spacefactor3000\relax}%
\providecommand \BibitemShut  [1]{\csname bibitem#1\endcsname}%
\let\auto@bib@innerbib\@empty
\bibitem [{\citenamefont {Hahn}\ and\ \citenamefont
  {Strassmann}(1939)}]{Hahn1939}%
  \BibitemOpen
  \bibfield  {author} {\bibinfo {author} {\bibfnamefont {O.}~\bibnamefont
  {Hahn}}\ and\ \bibinfo {author} {\bibfnamefont {F.}~\bibnamefont
  {Strassmann}},\ }\href {\doibase 10.1007/BF01488241} {\bibfield  {journal}
  {\bibinfo  {journal} {Naturwissenschaften}\ }\textbf {\bibinfo {volume}
  {27}},\ \bibinfo {pages} {11} (\bibinfo {year} {1939})}\BibitemShut {NoStop}%
\bibitem [{\citenamefont {Meitner}\ and\ \citenamefont
  {Frisch}(1939)}]{Meitner1939}%
  \BibitemOpen
  \bibfield  {author} {\bibinfo {author} {\bibfnamefont {L.}~\bibnamefont
  {Meitner}}\ and\ \bibinfo {author} {\bibfnamefont {O.~R.}\ \bibnamefont
  {Frisch}},\ }\href {\doibase 10.1038/143239a0} {\bibfield  {journal}
  {\bibinfo  {journal} {Nature}\ }\textbf {\bibinfo {volume} {143}},\ \bibinfo
  {pages} {239} (\bibinfo {year} {1939})}\BibitemShut {NoStop}%
\bibitem [{\citenamefont {Andreyev}\ \emph {et~al.}(2018)\citenamefont
  {Andreyev}, \citenamefont {Nishio},\ and\ \citenamefont
  {Schmidt}}]{Andreyev2018-ks}%
  \BibitemOpen
  \bibfield  {author} {\bibinfo {author} {\bibfnamefont {A.~N.}\ \bibnamefont
  {Andreyev}}, \bibinfo {author} {\bibfnamefont {K.}~\bibnamefont {Nishio}}, \
  and\ \bibinfo {author} {\bibfnamefont {K.-H.}\ \bibnamefont {Schmidt}},\
  }\href@noop {} {\bibfield  {journal} {\bibinfo  {journal} {Rep Prog Phys}\
  }\textbf {\bibinfo {volume} {81}},\ \bibinfo {pages} {016301} (\bibinfo
  {year} {2018})}\BibitemShut {NoStop}%
\bibitem [{\citenamefont {Schmidt}\ and\ \citenamefont
  {Jurado}(2018)}]{Schmidt_2018}%
  \BibitemOpen
  \bibfield  {author} {\bibinfo {author} {\bibfnamefont {K.-H.}\ \bibnamefont
  {Schmidt}}\ and\ \bibinfo {author} {\bibfnamefont {B.}~\bibnamefont
  {Jurado}},\ }\href {\doibase 10.1088/1361-6633/aacfa7} {\bibfield  {journal}
  {\bibinfo  {journal} {Reports on Progress in Physics}\ }\textbf {\bibinfo
  {volume} {81}},\ \bibinfo {pages} {106301} (\bibinfo {year}
  {2018})}\BibitemShut {NoStop}%
\bibitem [{\citenamefont {Rimpault}\ \emph {et~al.}(2012)\citenamefont
  {Rimpault}, \citenamefont {Bernard}, \citenamefont {Blanchet}, \citenamefont
  {Vaglio-Gaudard}, \citenamefont {Ravaux},\ and\ \citenamefont
  {Santamarina}}]{RIMPAULT20123}%
  \BibitemOpen
  \bibfield  {author} {\bibinfo {author} {\bibfnamefont {G.}~\bibnamefont
  {Rimpault}}, \bibinfo {author} {\bibfnamefont {D.}~\bibnamefont {Bernard}},
  \bibinfo {author} {\bibfnamefont {D.}~\bibnamefont {Blanchet}}, \bibinfo
  {author} {\bibfnamefont {C.}~\bibnamefont {Vaglio-Gaudard}}, \bibinfo
  {author} {\bibfnamefont {S.}~\bibnamefont {Ravaux}}, \ and\ \bibinfo {author}
  {\bibfnamefont {A.}~\bibnamefont {Santamarina}},\ }\href {\doibase
  https://doi.org/10.1016/j.phpro.2012.04.002} {\bibfield  {journal} {\bibinfo
  {journal} {Physics Procedia}\ }\textbf {\bibinfo {volume} {31}},\ \bibinfo
  {pages} {3} (\bibinfo {year} {2012})},\ \bibinfo {note} {gAMMA-1 Emission of
  Prompt Gamma-Rays in Fission and Related Topics}\BibitemShut {NoStop}%
\bibitem [{\citenamefont {Lemaire}\ \emph {et~al.}(2015)\citenamefont
  {Lemaire}, \citenamefont {Vaglio-Gaudard}, \citenamefont {Lyoussi},\ and\
  \citenamefont {Reynard-Carette}}]{Lemaire_2015}%
  \BibitemOpen
  \bibfield  {author} {\bibinfo {author} {\bibfnamefont {M.}~\bibnamefont
  {Lemaire}}, \bibinfo {author} {\bibfnamefont {C.}~\bibnamefont
  {Vaglio-Gaudard}}, \bibinfo {author} {\bibfnamefont {A.}~\bibnamefont
  {Lyoussi}}, \ and\ \bibinfo {author} {\bibfnamefont {C.}~\bibnamefont
  {Reynard-Carette}},\ }\href {\doibase 10.1080/00223131.2015.1009957}
  {\bibfield  {journal} {\bibinfo  {journal} {Journal of Nuclear Science and
  Technology}\ }\textbf {\bibinfo {volume} {52}},\ \bibinfo {pages} {1093}
  (\bibinfo {year} {2015})},\ \Eprint
  {http://arxiv.org/abs/https://doi.org/10.1080/00223131.2015.1009957}
  {https://doi.org/10.1080/00223131.2015.1009957} \BibitemShut {NoStop}%
\bibitem [{\citenamefont {Pleasonton}\ \emph {et~al.}(1972)\citenamefont
  {Pleasonton}, \citenamefont {Ferguson},\ and\ \citenamefont
  {Schmitt}}]{Pleasonton_PhysRevC.6.1023}%
  \BibitemOpen
  \bibfield  {author} {\bibinfo {author} {\bibfnamefont {F.}~\bibnamefont
  {Pleasonton}}, \bibinfo {author} {\bibfnamefont {R.~L.}\ \bibnamefont
  {Ferguson}}, \ and\ \bibinfo {author} {\bibfnamefont {H.~W.}\ \bibnamefont
  {Schmitt}},\ }\href {\doibase 10.1103/PhysRevC.6.1023} {\bibfield  {journal}
  {\bibinfo  {journal} {Phys. Rev. C}\ }\textbf {\bibinfo {volume} {6}},\
  \bibinfo {pages} {1023} (\bibinfo {year} {1972})}\BibitemShut {NoStop}%
\bibitem [{\citenamefont {Chebboubi}\ \emph {et~al.}(2021)\citenamefont
  {Chebboubi}, \citenamefont {Kessedjian}, \citenamefont {Serot}, \citenamefont
  {Faust}, \citenamefont {K{\"o}ster}, \citenamefont {Litaize}, \citenamefont
  {Sage}, \citenamefont {Blanc}, \citenamefont {Bernard}, \citenamefont
  {Letourneau}, \citenamefont {Materna}, \citenamefont {M{\'e}plan},
  \citenamefont {Mutti}, \citenamefont {Rapala},\ and\ \citenamefont
  {Ramdhane}}]{Chebboubi2021}%
  \BibitemOpen
  \bibfield  {author} {\bibinfo {author} {\bibfnamefont {A.}~\bibnamefont
  {Chebboubi}}, \bibinfo {author} {\bibfnamefont {G.}~\bibnamefont
  {Kessedjian}}, \bibinfo {author} {\bibfnamefont {O.}~\bibnamefont {Serot}},
  \bibinfo {author} {\bibfnamefont {H.}~\bibnamefont {Faust}}, \bibinfo
  {author} {\bibfnamefont {U.}~\bibnamefont {K{\"o}ster}}, \bibinfo {author}
  {\bibfnamefont {O.}~\bibnamefont {Litaize}}, \bibinfo {author} {\bibfnamefont
  {C.}~\bibnamefont {Sage}}, \bibinfo {author} {\bibfnamefont {A.}~\bibnamefont
  {Blanc}}, \bibinfo {author} {\bibfnamefont {D.}~\bibnamefont {Bernard}},
  \bibinfo {author} {\bibfnamefont {A.}~\bibnamefont {Letourneau}}, \bibinfo
  {author} {\bibfnamefont {T.}~\bibnamefont {Materna}}, \bibinfo {author}
  {\bibfnamefont {O.}~\bibnamefont {M{\'e}plan}}, \bibinfo {author}
  {\bibfnamefont {P.}~\bibnamefont {Mutti}}, \bibinfo {author} {\bibfnamefont
  {M.}~\bibnamefont {Rapala}}, \ and\ \bibinfo {author} {\bibfnamefont
  {M.}~\bibnamefont {Ramdhane}},\ }\href
  {https://doi.org/10.1140/epja/s10050-021-00645-y} {\bibfield  {journal}
  {\bibinfo  {journal} {The European Physical Journal A}\ }\textbf {\bibinfo
  {volume} {57}},\ \bibinfo {pages} {335} (\bibinfo {year} {2021})}\BibitemShut
  {NoStop}%
\bibitem [{\citenamefont {Armbruster}\ \emph {et~al.}(1976)\citenamefont
  {Armbruster}, \citenamefont {Asghar}, \citenamefont {Bocquet}, \citenamefont
  {Decker}, \citenamefont {Ewald}, \citenamefont {Greif}, \citenamefont {Moll},
  \citenamefont {Pfeiffer}, \citenamefont {Schrader}, \citenamefont
  {Schussler}, \citenamefont {Siegert},\ and\ \citenamefont
  {Wollnik}}]{ARMBRUSTER1976213}%
  \BibitemOpen
  \bibfield  {author} {\bibinfo {author} {\bibfnamefont {P.}~\bibnamefont
  {Armbruster}}, \bibinfo {author} {\bibfnamefont {M.}~\bibnamefont {Asghar}},
  \bibinfo {author} {\bibfnamefont {J.}~\bibnamefont {Bocquet}}, \bibinfo
  {author} {\bibfnamefont {R.}~\bibnamefont {Decker}}, \bibinfo {author}
  {\bibfnamefont {H.}~\bibnamefont {Ewald}}, \bibinfo {author} {\bibfnamefont
  {J.}~\bibnamefont {Greif}}, \bibinfo {author} {\bibfnamefont
  {E.}~\bibnamefont {Moll}}, \bibinfo {author} {\bibfnamefont {B.}~\bibnamefont
  {Pfeiffer}}, \bibinfo {author} {\bibfnamefont {H.}~\bibnamefont {Schrader}},
  \bibinfo {author} {\bibfnamefont {F.}~\bibnamefont {Schussler}}, \bibinfo
  {author} {\bibfnamefont {G.}~\bibnamefont {Siegert}}, \ and\ \bibinfo
  {author} {\bibfnamefont {H.}~\bibnamefont {Wollnik}},\ }\href {\doibase
  https://doi.org/10.1016/0029-554X(76)90677-7} {\bibfield  {journal} {\bibinfo
   {journal} {Nuclear Instruments and Methods}\ }\textbf {\bibinfo {volume}
  {139}},\ \bibinfo {pages} {213} (\bibinfo {year} {1976})}\BibitemShut
  {NoStop}%
\bibitem [{\citenamefont {Verbinski}\ \emph {et~al.}(1973)\citenamefont
  {Verbinski}, \citenamefont {Weber},\ and\ \citenamefont
  {Sund}}]{Verbinski_PhysRevC.7.1173}%
  \BibitemOpen
  \bibfield  {author} {\bibinfo {author} {\bibfnamefont {V.~V.}\ \bibnamefont
  {Verbinski}}, \bibinfo {author} {\bibfnamefont {H.}~\bibnamefont {Weber}}, \
  and\ \bibinfo {author} {\bibfnamefont {R.~E.}\ \bibnamefont {Sund}},\ }\href
  {\doibase 10.1103/PhysRevC.7.1173} {\bibfield  {journal} {\bibinfo  {journal}
  {Phys. Rev. C}\ }\textbf {\bibinfo {volume} {7}},\ \bibinfo {pages} {1173}
  (\bibinfo {year} {1973})}\BibitemShut {NoStop}%
\bibitem [{\citenamefont {Chyzh}\ \emph {et~al.}(2012)\citenamefont {Chyzh},
  \citenamefont {Wu}, \citenamefont {Kwan}, \citenamefont {Henderson},
  \citenamefont {Gostic}, \citenamefont {Bredeweg}, \citenamefont {Haight},
  \citenamefont {Hayes-Sterbenz}, \citenamefont {Jandel}, \citenamefont
  {O'Donnell},\ and\ \citenamefont {Ullmann}}]{Chyzh_PhysRevC.85.021601}%
  \BibitemOpen
  \bibfield  {author} {\bibinfo {author} {\bibfnamefont {A.}~\bibnamefont
  {Chyzh}}, \bibinfo {author} {\bibfnamefont {C.~Y.}\ \bibnamefont {Wu}},
  \bibinfo {author} {\bibfnamefont {E.}~\bibnamefont {Kwan}}, \bibinfo {author}
  {\bibfnamefont {R.~A.}\ \bibnamefont {Henderson}}, \bibinfo {author}
  {\bibfnamefont {J.~M.}\ \bibnamefont {Gostic}}, \bibinfo {author}
  {\bibfnamefont {T.~A.}\ \bibnamefont {Bredeweg}}, \bibinfo {author}
  {\bibfnamefont {R.~C.}\ \bibnamefont {Haight}}, \bibinfo {author}
  {\bibfnamefont {A.~C.}\ \bibnamefont {Hayes-Sterbenz}}, \bibinfo {author}
  {\bibfnamefont {M.}~\bibnamefont {Jandel}}, \bibinfo {author} {\bibfnamefont
  {J.~M.}\ \bibnamefont {O'Donnell}}, \ and\ \bibinfo {author} {\bibfnamefont
  {J.~L.}\ \bibnamefont {Ullmann}},\ }\href
  {https://link.aps.org/doi/10.1103/PhysRevC.85.021601} {\bibfield  {journal}
  {\bibinfo  {journal} {Phys. Rev. C}\ }\textbf {\bibinfo {volume} {85}},\
  \bibinfo {pages} {021601} (\bibinfo {year} {2012})}\BibitemShut {NoStop}%
\bibitem [{\citenamefont {Heil}\ \emph {et~al.}(2001)\citenamefont {Heil},
  \citenamefont {Reifarth}, \citenamefont {Fowler}, \citenamefont {Haight},
  \citenamefont {K\"{a}ppeler}, \citenamefont {Rundberg}, \citenamefont
  {Seabury}, \citenamefont {Ullmann}, \citenamefont {Wilhelmy},\ and\
  \citenamefont {Wisshak}}]{HEIL2001229}%
  \BibitemOpen
  \bibfield  {author} {\bibinfo {author} {\bibfnamefont {M.}~\bibnamefont
  {Heil}}, \bibinfo {author} {\bibfnamefont {R.}~\bibnamefont {Reifarth}},
  \bibinfo {author} {\bibfnamefont {M.}~\bibnamefont {Fowler}}, \bibinfo
  {author} {\bibfnamefont {R.}~\bibnamefont {Haight}}, \bibinfo {author}
  {\bibfnamefont {F.}~\bibnamefont {K\"{a}ppeler}}, \bibinfo {author}
  {\bibfnamefont {R.}~\bibnamefont {Rundberg}}, \bibinfo {author}
  {\bibfnamefont {E.}~\bibnamefont {Seabury}}, \bibinfo {author} {\bibfnamefont
  {J.}~\bibnamefont {Ullmann}}, \bibinfo {author} {\bibfnamefont
  {J.}~\bibnamefont {Wilhelmy}}, \ and\ \bibinfo {author} {\bibfnamefont
  {K.}~\bibnamefont {Wisshak}},\ }\href {\doibase
  https://doi.org/10.1016/S0168-9002(00)00993-1} {\bibfield  {journal}
  {\bibinfo  {journal} {Nuclear Instruments and Methods in Physics Research
  Section A: Accelerators, Spectrometers, Detectors and Associated Equipment}\
  }\textbf {\bibinfo {volume} {459}},\ \bibinfo {pages} {229} (\bibinfo {year}
  {2001})}\BibitemShut {NoStop}%
\bibitem [{\citenamefont {Oberstedt}\ \emph {et~al.}(2015)\citenamefont
  {Oberstedt}, \citenamefont {Billnert}, \citenamefont {Hambsch},\ and\
  \citenamefont {Oberstedt}}]{Oberstedt_2015_PhysRevC.92.014618}%
  \BibitemOpen
  \bibfield  {author} {\bibinfo {author} {\bibfnamefont {A.}~\bibnamefont
  {Oberstedt}}, \bibinfo {author} {\bibfnamefont {R.}~\bibnamefont {Billnert}},
  \bibinfo {author} {\bibfnamefont {F.-J.}\ \bibnamefont {Hambsch}}, \ and\
  \bibinfo {author} {\bibfnamefont {S.}~\bibnamefont {Oberstedt}},\ }\href
  {\doibase 10.1103/PhysRevC.92.014618} {\bibfield  {journal} {\bibinfo
  {journal} {Phys. Rev. C}\ }\textbf {\bibinfo {volume} {92}},\ \bibinfo
  {pages} {014618} (\bibinfo {year} {2015})}\BibitemShut {NoStop}%
\bibitem [{\citenamefont {Oberstedt}\ \emph {et~al.}(2013)\citenamefont
  {Oberstedt}, \citenamefont {Borcea}, \citenamefont {Bry\`s}, \citenamefont
  {Gamboni}, \citenamefont {Geerts}, \citenamefont {Hambsch}, \citenamefont
  {Oberstedt},\ and\ \citenamefont {Vidali}}]{OBERSTEDT201331}%
  \BibitemOpen
  \bibfield  {author} {\bibinfo {author} {\bibfnamefont {S.}~\bibnamefont
  {Oberstedt}}, \bibinfo {author} {\bibfnamefont {R.}~\bibnamefont {Borcea}},
  \bibinfo {author} {\bibfnamefont {T.}~\bibnamefont {Bry\`s}}, \bibinfo
  {author} {\bibfnamefont {T.}~\bibnamefont {Gamboni}}, \bibinfo {author}
  {\bibfnamefont {W.}~\bibnamefont {Geerts}}, \bibinfo {author} {\bibfnamefont
  {F.-J.}\ \bibnamefont {Hambsch}}, \bibinfo {author} {\bibfnamefont
  {A.}~\bibnamefont {Oberstedt}}, \ and\ \bibinfo {author} {\bibfnamefont
  {M.}~\bibnamefont {Vidali}},\ }\href {\doibase
  https://doi.org/10.1016/j.nima.2013.02.029} {\bibfield  {journal} {\bibinfo
  {journal} {Nuclear Instruments and Methods in Physics Research Section A:
  Accelerators, Spectrometers, Detectors and Associated Equipment}\ }\textbf
  {\bibinfo {volume} {714}},\ \bibinfo {pages} {31} (\bibinfo {year}
  {2013})}\BibitemShut {NoStop}%
\bibitem [{\citenamefont {Wilhelmy}\ \emph {et~al.}(1972)\citenamefont
  {Wilhelmy}, \citenamefont {Cheifetz}, \citenamefont {Jared}, \citenamefont
  {Thompson}, \citenamefont {Bowman},\ and\ \citenamefont
  {Rasmussen}}]{Wilhelmy_PhysRevC.5.2041}%
  \BibitemOpen
  \bibfield  {author} {\bibinfo {author} {\bibfnamefont {J.~B.}\ \bibnamefont
  {Wilhelmy}}, \bibinfo {author} {\bibfnamefont {E.}~\bibnamefont {Cheifetz}},
  \bibinfo {author} {\bibfnamefont {R.~C.}\ \bibnamefont {Jared}}, \bibinfo
  {author} {\bibfnamefont {S.~G.}\ \bibnamefont {Thompson}}, \bibinfo {author}
  {\bibfnamefont {H.~R.}\ \bibnamefont {Bowman}}, \ and\ \bibinfo {author}
  {\bibfnamefont {J.~O.}\ \bibnamefont {Rasmussen}},\ }\href
  {https://link.aps.org/doi/10.1103/PhysRevC.5.2041} {\bibfield  {journal}
  {\bibinfo  {journal} {Phys. Rev. C}\ }\textbf {\bibinfo {volume} {5}},\
  \bibinfo {pages} {2041} (\bibinfo {year} {1972})}\BibitemShut {NoStop}%
\bibitem [{\citenamefont {Skarsv\aa{}g}(1980)}]{Skargvag_1980_PhysRevC.22.638}%
  \BibitemOpen
  \bibfield  {author} {\bibinfo {author} {\bibfnamefont {K.}~\bibnamefont
  {Skarsv\aa{}g}},\ }\href {\doibase 10.1103/PhysRevC.22.638} {\bibfield
  {journal} {\bibinfo  {journal} {Phys. Rev. C}\ }\textbf {\bibinfo {volume}
  {22}},\ \bibinfo {pages} {638} (\bibinfo {year} {1980})}\BibitemShut
  {NoStop}%
\bibitem [{\citenamefont {Wolf}\ and\ \citenamefont
  {Cheifetz}(1976)}]{Wolf_1976_PhysRevC.13.1952}%
  \BibitemOpen
  \bibfield  {author} {\bibinfo {author} {\bibfnamefont {A.}~\bibnamefont
  {Wolf}}\ and\ \bibinfo {author} {\bibfnamefont {E.}~\bibnamefont
  {Cheifetz}},\ }\href {\doibase 10.1103/PhysRevC.13.1952} {\bibfield
  {journal} {\bibinfo  {journal} {Phys. Rev. C}\ }\textbf {\bibinfo {volume}
  {13}},\ \bibinfo {pages} {1952} (\bibinfo {year} {1976})}\BibitemShut
  {NoStop}%
\bibitem [{\citenamefont {Nix}\ and\ \citenamefont
  {Swiatecki}(1965)}]{NIX19651}%
  \BibitemOpen
  \bibfield  {author} {\bibinfo {author} {\bibfnamefont {J.~R.}\ \bibnamefont
  {Nix}}\ and\ \bibinfo {author} {\bibfnamefont {W.~J.}\ \bibnamefont
  {Swiatecki}},\ }\href {\doibase https://doi.org/10.1016/0029-5582(65)90038-6}
  {\bibfield  {journal} {\bibinfo  {journal} {Nuclear Physics}\ }\textbf
  {\bibinfo {volume} {71}},\ \bibinfo {pages} {1} (\bibinfo {year}
  {1965})}\BibitemShut {NoStop}%
\bibitem [{\citenamefont {Hoffman}(1964)}]{Hoffman_PhysRev.133.B714}%
  \BibitemOpen
  \bibfield  {author} {\bibinfo {author} {\bibfnamefont {M.~M.}\ \bibnamefont
  {Hoffman}},\ }\href {\doibase 10.1103/PhysRev.133.B714} {\bibfield  {journal}
  {\bibinfo  {journal} {Phys. Rev.}\ }\textbf {\bibinfo {volume} {133}},\
  \bibinfo {pages} {B714} (\bibinfo {year} {1964})}\BibitemShut {NoStop}%
\bibitem [{\citenamefont {Smith}\ \emph {et~al.}(1999)\citenamefont {Smith},
  \citenamefont {Simpson}, \citenamefont {Billowes}, \citenamefont {Dagnall},
  \citenamefont {Durell}, \citenamefont {Freeman}, \citenamefont {Leddy},
  \citenamefont {Phillips}, \citenamefont {Roach}, \citenamefont {Smith},
  \citenamefont {Jungclaus}, \citenamefont {Lieb}, \citenamefont {Teich},
  \citenamefont {Gall}, \citenamefont {Hoellinger}, \citenamefont {Schulz},
  \citenamefont {Ahmad}, \citenamefont {Greene},\ and\ \citenamefont
  {Algora}}]{Smith_PhysRevC.60.064611}%
  \BibitemOpen
  \bibfield  {author} {\bibinfo {author} {\bibfnamefont {A.~G.}\ \bibnamefont
  {Smith}}, \bibinfo {author} {\bibfnamefont {G.~S.}\ \bibnamefont {Simpson}},
  \bibinfo {author} {\bibfnamefont {J.}~\bibnamefont {Billowes}}, \bibinfo
  {author} {\bibfnamefont {P.~J.}\ \bibnamefont {Dagnall}}, \bibinfo {author}
  {\bibfnamefont {J.~L.}\ \bibnamefont {Durell}}, \bibinfo {author}
  {\bibfnamefont {S.~J.}\ \bibnamefont {Freeman}}, \bibinfo {author}
  {\bibfnamefont {M.}~\bibnamefont {Leddy}}, \bibinfo {author} {\bibfnamefont
  {W.~R.}\ \bibnamefont {Phillips}}, \bibinfo {author} {\bibfnamefont {A.~A.}\
  \bibnamefont {Roach}}, \bibinfo {author} {\bibfnamefont {J.~F.}\ \bibnamefont
  {Smith}}, \bibinfo {author} {\bibfnamefont {A.}~\bibnamefont {Jungclaus}},
  \bibinfo {author} {\bibfnamefont {K.~P.}\ \bibnamefont {Lieb}}, \bibinfo
  {author} {\bibfnamefont {C.}~\bibnamefont {Teich}}, \bibinfo {author}
  {\bibfnamefont {B.~J.~P.}\ \bibnamefont {Gall}}, \bibinfo {author}
  {\bibfnamefont {F.}~\bibnamefont {Hoellinger}}, \bibinfo {author}
  {\bibfnamefont {N.}~\bibnamefont {Schulz}}, \bibinfo {author} {\bibfnamefont
  {I.}~\bibnamefont {Ahmad}}, \bibinfo {author} {\bibfnamefont {J.~P.}\
  \bibnamefont {Greene}}, \ and\ \bibinfo {author} {\bibfnamefont
  {A.}~\bibnamefont {Algora}},\ }\href {\doibase 10.1103/PhysRevC.60.064611}
  {\bibfield  {journal} {\bibinfo  {journal} {Phys. Rev. C}\ }\textbf {\bibinfo
  {volume} {60}},\ \bibinfo {pages} {064611} (\bibinfo {year}
  {1999})}\BibitemShut {NoStop}%
\bibitem [{\citenamefont {Wilson}\ \emph {et~al.}(2021)\citenamefont {Wilson},
  \citenamefont {Thisse}, \citenamefont {Lebois}, \citenamefont
  {Jovan{\v{c}}evi{\'{c}}}, \citenamefont {Gjestvang}, \citenamefont {Canavan},
  \citenamefont {Rudigier}, \citenamefont {{\'E}tasse}, \citenamefont {Gerst},
  \citenamefont {Gaudefroy}, \citenamefont {Adamska}, \citenamefont {Adsley},
  \citenamefont {Algora}, \citenamefont {Babo}, \citenamefont {Belvedere},
  \citenamefont {Benito}, \citenamefont {Benzoni}, \citenamefont {Blazhev},
  \citenamefont {Boso}, \citenamefont {Bottoni}, \citenamefont {Bunce},
  \citenamefont {Chakma}, \citenamefont {Cieplicka-Ory{\'{n}}czak},
  \citenamefont {Courtin}, \citenamefont {Cort{\'e}s}, \citenamefont {Davies},
  \citenamefont {Delafosse}, \citenamefont {Fallot}, \citenamefont {Fornal},
  \citenamefont {Fraile}, \citenamefont {Gottardo}, \citenamefont {Guadilla},
  \citenamefont {H{\"a}fner}, \citenamefont {Hauschild}, \citenamefont {Heine},
  \citenamefont {Henrich}, \citenamefont {Homm}, \citenamefont {Ibrahim},
  \citenamefont {Iskra}, \citenamefont {Ivanov}, \citenamefont {Jazrawi},
  \citenamefont {Korgul}, \citenamefont {Koseoglou}, \citenamefont {Kr{\"o}ll},
  \citenamefont {Kurtukian-Nieto}, \citenamefont {Le~Meur}, \citenamefont
  {Leoni}, \citenamefont {Ljungvall}, \citenamefont {Lopez-Martens},
  \citenamefont {Lozeva}, \citenamefont {Matea}, \citenamefont {Miernik},
  \citenamefont {Nemer}, \citenamefont {Oberstedt}, \citenamefont {Paulsen},
  \citenamefont {Piersa}, \citenamefont {Popovitch}, \citenamefont {Porzio},
  \citenamefont {Qi}, \citenamefont {Ralet}, \citenamefont {Regan},
  \citenamefont {Rezynkina}, \citenamefont {S{\'a}nchez-Tembleque},
  \citenamefont {Siem}, \citenamefont {Schmitt}, \citenamefont
  {S{\"o}derstr{\"o}m}, \citenamefont {S{\"u}rder}, \citenamefont {Tocabens},
  \citenamefont {Vedia}, \citenamefont {Verney}, \citenamefont {Warr},
  \citenamefont {Wasilewska}, \citenamefont {Wiederhold}, \citenamefont
  {Yavahchova}, \citenamefont {Zeiser},\ and\ \citenamefont
  {Ziliani}}]{Wilson2021}%
  \BibitemOpen
  \bibfield  {author} {\bibinfo {author} {\bibfnamefont {J.~N.}\ \bibnamefont
  {Wilson}}, \bibinfo {author} {\bibfnamefont {D.}~\bibnamefont {Thisse}},
  \bibinfo {author} {\bibfnamefont {M.}~\bibnamefont {Lebois}}, \bibinfo
  {author} {\bibfnamefont {N.}~\bibnamefont {Jovan{\v{c}}evi{\'{c}}}}, \bibinfo
  {author} {\bibfnamefont {D.}~\bibnamefont {Gjestvang}}, \bibinfo {author}
  {\bibfnamefont {R.}~\bibnamefont {Canavan}}, \bibinfo {author} {\bibfnamefont
  {M.}~\bibnamefont {Rudigier}}, \bibinfo {author} {\bibfnamefont
  {D.}~\bibnamefont {{\'E}tasse}}, \bibinfo {author} {\bibfnamefont {R.-B.}\
  \bibnamefont {Gerst}}, \bibinfo {author} {\bibfnamefont {L.}~\bibnamefont
  {Gaudefroy}}, \bibinfo {author} {\bibfnamefont {E.}~\bibnamefont {Adamska}},
  \bibinfo {author} {\bibfnamefont {P.}~\bibnamefont {Adsley}}, \bibinfo
  {author} {\bibfnamefont {A.}~\bibnamefont {Algora}}, \bibinfo {author}
  {\bibfnamefont {M.}~\bibnamefont {Babo}}, \bibinfo {author} {\bibfnamefont
  {K.}~\bibnamefont {Belvedere}}, \bibinfo {author} {\bibfnamefont
  {J.}~\bibnamefont {Benito}}, \bibinfo {author} {\bibfnamefont
  {G.}~\bibnamefont {Benzoni}}, \bibinfo {author} {\bibfnamefont
  {A.}~\bibnamefont {Blazhev}}, \bibinfo {author} {\bibfnamefont
  {A.}~\bibnamefont {Boso}}, \bibinfo {author} {\bibfnamefont {S.}~\bibnamefont
  {Bottoni}}, \bibinfo {author} {\bibfnamefont {M.}~\bibnamefont {Bunce}},
  \bibinfo {author} {\bibfnamefont {R.}~\bibnamefont {Chakma}}, \bibinfo
  {author} {\bibfnamefont {N.}~\bibnamefont {Cieplicka-Ory{\'{n}}czak}},
  \bibinfo {author} {\bibfnamefont {S.}~\bibnamefont {Courtin}}, \bibinfo
  {author} {\bibfnamefont {M.~L.}\ \bibnamefont {Cort{\'e}s}}, \bibinfo
  {author} {\bibfnamefont {P.}~\bibnamefont {Davies}}, \bibinfo {author}
  {\bibfnamefont {C.}~\bibnamefont {Delafosse}}, \bibinfo {author}
  {\bibfnamefont {M.}~\bibnamefont {Fallot}}, \bibinfo {author} {\bibfnamefont
  {B.}~\bibnamefont {Fornal}}, \bibinfo {author} {\bibfnamefont
  {L.}~\bibnamefont {Fraile}}, \bibinfo {author} {\bibfnamefont
  {A.}~\bibnamefont {Gottardo}}, \bibinfo {author} {\bibfnamefont
  {V.}~\bibnamefont {Guadilla}}, \bibinfo {author} {\bibfnamefont
  {G.}~\bibnamefont {H{\"a}fner}}, \bibinfo {author} {\bibfnamefont
  {K.}~\bibnamefont {Hauschild}}, \bibinfo {author} {\bibfnamefont
  {M.}~\bibnamefont {Heine}}, \bibinfo {author} {\bibfnamefont
  {C.}~\bibnamefont {Henrich}}, \bibinfo {author} {\bibfnamefont
  {I.}~\bibnamefont {Homm}}, \bibinfo {author} {\bibfnamefont {F.}~\bibnamefont
  {Ibrahim}}, \bibinfo {author} {\bibfnamefont {{\L}.~W.}\ \bibnamefont
  {Iskra}}, \bibinfo {author} {\bibfnamefont {P.}~\bibnamefont {Ivanov}},
  \bibinfo {author} {\bibfnamefont {S.}~\bibnamefont {Jazrawi}}, \bibinfo
  {author} {\bibfnamefont {A.}~\bibnamefont {Korgul}}, \bibinfo {author}
  {\bibfnamefont {P.}~\bibnamefont {Koseoglou}}, \bibinfo {author}
  {\bibfnamefont {T.}~\bibnamefont {Kr{\"o}ll}}, \bibinfo {author}
  {\bibfnamefont {T.}~\bibnamefont {Kurtukian-Nieto}}, \bibinfo {author}
  {\bibfnamefont {L.}~\bibnamefont {Le~Meur}}, \bibinfo {author} {\bibfnamefont
  {S.}~\bibnamefont {Leoni}}, \bibinfo {author} {\bibfnamefont
  {J.}~\bibnamefont {Ljungvall}}, \bibinfo {author} {\bibfnamefont
  {A.}~\bibnamefont {Lopez-Martens}}, \bibinfo {author} {\bibfnamefont
  {R.}~\bibnamefont {Lozeva}}, \bibinfo {author} {\bibfnamefont
  {I.}~\bibnamefont {Matea}}, \bibinfo {author} {\bibfnamefont
  {K.}~\bibnamefont {Miernik}}, \bibinfo {author} {\bibfnamefont
  {J.}~\bibnamefont {Nemer}}, \bibinfo {author} {\bibfnamefont
  {S.}~\bibnamefont {Oberstedt}}, \bibinfo {author} {\bibfnamefont
  {W.}~\bibnamefont {Paulsen}}, \bibinfo {author} {\bibfnamefont
  {M.}~\bibnamefont {Piersa}}, \bibinfo {author} {\bibfnamefont
  {Y.}~\bibnamefont {Popovitch}}, \bibinfo {author} {\bibfnamefont
  {C.}~\bibnamefont {Porzio}}, \bibinfo {author} {\bibfnamefont
  {L.}~\bibnamefont {Qi}}, \bibinfo {author} {\bibfnamefont {D.}~\bibnamefont
  {Ralet}}, \bibinfo {author} {\bibfnamefont {P.~H.}\ \bibnamefont {Regan}},
  \bibinfo {author} {\bibfnamefont {K.}~\bibnamefont {Rezynkina}}, \bibinfo
  {author} {\bibfnamefont {V.}~\bibnamefont {S{\'a}nchez-Tembleque}}, \bibinfo
  {author} {\bibfnamefont {S.}~\bibnamefont {Siem}}, \bibinfo {author}
  {\bibfnamefont {C.}~\bibnamefont {Schmitt}}, \bibinfo {author} {\bibfnamefont
  {P.-A.}\ \bibnamefont {S{\"o}derstr{\"o}m}}, \bibinfo {author} {\bibfnamefont
  {C.}~\bibnamefont {S{\"u}rder}}, \bibinfo {author} {\bibfnamefont
  {G.}~\bibnamefont {Tocabens}}, \bibinfo {author} {\bibfnamefont
  {V.}~\bibnamefont {Vedia}}, \bibinfo {author} {\bibfnamefont
  {D.}~\bibnamefont {Verney}}, \bibinfo {author} {\bibfnamefont
  {N.}~\bibnamefont {Warr}}, \bibinfo {author} {\bibfnamefont {B.}~\bibnamefont
  {Wasilewska}}, \bibinfo {author} {\bibfnamefont {J.}~\bibnamefont
  {Wiederhold}}, \bibinfo {author} {\bibfnamefont {M.}~\bibnamefont
  {Yavahchova}}, \bibinfo {author} {\bibfnamefont {F.}~\bibnamefont {Zeiser}},
  \ and\ \bibinfo {author} {\bibfnamefont {S.}~\bibnamefont {Ziliani}},\ }\href
  {\doibase 10.1038/s41586-021-03304-w} {\bibfield  {journal} {\bibinfo
  {journal} {Nature}\ }\textbf {\bibinfo {volume} {590}},\ \bibinfo {pages}
  {566} (\bibinfo {year} {2021})}\BibitemShut {NoStop}%
\bibitem [{\citenamefont {Randrup}\ and\ \citenamefont
  {Vogt}(2021)}]{Randrup_2021_PhysRevLett.127.062502}%
  \BibitemOpen
  \bibfield  {author} {\bibinfo {author} {\bibfnamefont {J.}~\bibnamefont
  {Randrup}}\ and\ \bibinfo {author} {\bibfnamefont {R.}~\bibnamefont {Vogt}},\
  }\href {\doibase 10.1103/PhysRevLett.127.062502} {\bibfield  {journal}
  {\bibinfo  {journal} {Phys. Rev. Lett.}\ }\textbf {\bibinfo {volume} {127}},\
  \bibinfo {pages} {062502} (\bibinfo {year} {2021})}\BibitemShut {NoStop}%
\bibitem [{\citenamefont {Bertsch}\ \emph {et~al.}(2019)\citenamefont
  {Bertsch}, \citenamefont {Kawano},\ and\ \citenamefont
  {Robledo}}]{Bertsch_2019_PhysRevC.99.034603}%
  \BibitemOpen
  \bibfield  {author} {\bibinfo {author} {\bibfnamefont {G.~F.}\ \bibnamefont
  {Bertsch}}, \bibinfo {author} {\bibfnamefont {T.}~\bibnamefont {Kawano}}, \
  and\ \bibinfo {author} {\bibfnamefont {L.~M.}\ \bibnamefont {Robledo}},\
  }\href {\doibase 10.1103/PhysRevC.99.034603} {\bibfield  {journal} {\bibinfo
  {journal} {Phys. Rev. C}\ }\textbf {\bibinfo {volume} {99}},\ \bibinfo
  {pages} {034603} (\bibinfo {year} {2019})}\BibitemShut {NoStop}%
\bibitem [{\citenamefont {Bertsch}(2020)}]{Bertsch_2020_PhysRevC.101.034617}%
  \BibitemOpen
  \bibfield  {author} {\bibinfo {author} {\bibfnamefont {G.~F.}\ \bibnamefont
  {Bertsch}},\ }\href {\doibase 10.1103/PhysRevC.101.034617} {\bibfield
  {journal} {\bibinfo  {journal} {Phys. Rev. C}\ }\textbf {\bibinfo {volume}
  {101}},\ \bibinfo {pages} {034617} (\bibinfo {year} {2020})}\BibitemShut
  {NoStop}%
\bibitem [{\citenamefont {Bertsch}\ and\ \citenamefont
  {Hagino}(2023)}]{Bertsch_2023_PhysRevC.107.044615}%
  \BibitemOpen
  \bibfield  {author} {\bibinfo {author} {\bibfnamefont {G.~F.}\ \bibnamefont
  {Bertsch}}\ and\ \bibinfo {author} {\bibfnamefont {K.}~\bibnamefont
  {Hagino}},\ }\href {\doibase 10.1103/PhysRevC.107.044615} {\bibfield
  {journal} {\bibinfo  {journal} {Phys. Rev. C}\ }\textbf {\bibinfo {volume}
  {107}},\ \bibinfo {pages} {044615} (\bibinfo {year} {2023})}\BibitemShut
  {NoStop}%
\bibitem [{\citenamefont {Randrup}\ \emph {et~al.}(2022)\citenamefont
  {Randrup}, \citenamefont {D\o{}ssing},\ and\ \citenamefont
  {Vogt}}]{Randrup_2022_PhysRevC.106.014609}%
  \BibitemOpen
  \bibfield  {author} {\bibinfo {author} {\bibfnamefont {J.}~\bibnamefont
  {Randrup}}, \bibinfo {author} {\bibfnamefont {T.}~\bibnamefont {D\o{}ssing}},
  \ and\ \bibinfo {author} {\bibfnamefont {R.}~\bibnamefont {Vogt}},\ }\href
  {\doibase 10.1103/PhysRevC.106.014609} {\bibfield  {journal} {\bibinfo
  {journal} {Phys. Rev. C}\ }\textbf {\bibinfo {volume} {106}},\ \bibinfo
  {pages} {014609} (\bibinfo {year} {2022})}\BibitemShut {NoStop}%
\bibitem [{\citenamefont {Bulgac}\ \emph {et~al.}(2022)\citenamefont {Bulgac},
  \citenamefont {Abdurrahman}, \citenamefont {Godbey},\ and\ \citenamefont
  {Stetcu}}]{Bulgac_2022_PhysRevLett.128.022501}%
  \BibitemOpen
  \bibfield  {author} {\bibinfo {author} {\bibfnamefont {A.}~\bibnamefont
  {Bulgac}}, \bibinfo {author} {\bibfnamefont {I.}~\bibnamefont {Abdurrahman}},
  \bibinfo {author} {\bibfnamefont {K.}~\bibnamefont {Godbey}}, \ and\ \bibinfo
  {author} {\bibfnamefont {I.}~\bibnamefont {Stetcu}},\ }\href
  {https://link.aps.org/doi/10.1103/PhysRevLett.128.022501} {\bibfield
  {journal} {\bibinfo  {journal} {Phys. Rev. Lett.}\ }\textbf {\bibinfo
  {volume} {128}},\ \bibinfo {pages} {022501} (\bibinfo {year}
  {2022})}\BibitemShut {NoStop}%
\bibitem [{\citenamefont {Litaize}\ \emph {et~al.}(2015)\citenamefont
  {Litaize}, \citenamefont {Serot},\ and\ \citenamefont {Berge}}]{Litaize2015}%
  \BibitemOpen
  \bibfield  {author} {\bibinfo {author} {\bibfnamefont {O.}~\bibnamefont
  {Litaize}}, \bibinfo {author} {\bibfnamefont {O.}~\bibnamefont {Serot}}, \
  and\ \bibinfo {author} {\bibfnamefont {L.}~\bibnamefont {Berge}},\ }\href
  {\doibase https://doi.org/10.1140/epja/i2015-15177-9} {\bibfield  {journal}
  {\bibinfo  {journal} {The European Physics Journal A}\ }\textbf {\bibinfo
  {volume} {51}},\ \bibinfo {pages} {117} (\bibinfo {year} {2015})}\BibitemShut
  {NoStop}%
\bibitem [{\citenamefont {{Oberstedt, Andreas}}\ \emph
  {et~al.}(2018)\citenamefont {{Oberstedt, Andreas}}, \citenamefont {{Billnert,
  Robert}}, \citenamefont {{Gatera, Ang\'elique}}, \citenamefont {{G\"o\"ok,
  Alf}},\ and\ \citenamefont {{Oberstedt, Stephan}}}]{oberstedt_2018}%
  \BibitemOpen
  \bibfield  {author} {\bibinfo {author} {\bibnamefont {{Oberstedt, Andreas}}},
  \bibinfo {author} {\bibnamefont {{Billnert, Robert}}}, \bibinfo {author}
  {\bibnamefont {{Gatera, Ang\'elique}}}, \bibinfo {author} {\bibnamefont
  {{G\"o\"ok, Alf}}}, \ and\ \bibinfo {author} {\bibnamefont {{Oberstedt,
  Stephan}}},\ }\href {\doibase 10.1051/epjconf/201819303005} {\bibfield
  {journal} {\bibinfo  {journal} {EPJ Web Conf.}\ }\textbf {\bibinfo {volume}
  {193}},\ \bibinfo {pages} {03005} (\bibinfo {year} {2018})}\BibitemShut
  {NoStop}%
\bibitem [{\citenamefont {Stuchbery}(2003)}]{STUCHBERY200369}%
  \BibitemOpen
  \bibfield  {author} {\bibinfo {author} {\bibfnamefont {A.~E.}\ \bibnamefont
  {Stuchbery}},\ }\href {\doibase
  https://doi.org/10.1016/S0375-9474(03)01157-6} {\bibfield  {journal}
  {\bibinfo  {journal} {Nuclear Physics A}\ }\textbf {\bibinfo {volume}
  {723}},\ \bibinfo {pages} {69} (\bibinfo {year} {2003})}\BibitemShut
  {NoStop}%
\bibitem [{\citenamefont {Olliver}\ \emph {et~al.}(2003)\citenamefont
  {Olliver}, \citenamefont {Glasmacher},\ and\ \citenamefont
  {Stuchbery}}]{Heather_2003_PhysRevC.68.044312}%
  \BibitemOpen
  \bibfield  {author} {\bibinfo {author} {\bibfnamefont {H.}~\bibnamefont
  {Olliver}}, \bibinfo {author} {\bibfnamefont {T.}~\bibnamefont {Glasmacher}},
  \ and\ \bibinfo {author} {\bibfnamefont {A.~E.}\ \bibnamefont {Stuchbery}},\
  }\href {\doibase 10.1103/PhysRevC.68.044312} {\bibfield  {journal} {\bibinfo
  {journal} {Phys. Rev. C}\ }\textbf {\bibinfo {volume} {68}},\ \bibinfo
  {pages} {044312} (\bibinfo {year} {2003})}\BibitemShut {NoStop}%
\bibitem [{\citenamefont {Diamond}\ \emph {et~al.}(1966)\citenamefont
  {Diamond}, \citenamefont {Matthias}, \citenamefont {Newton},\ and\
  \citenamefont {Stephens}}]{Diamond_1966_PhysRevLett.16.1205}%
  \BibitemOpen
  \bibfield  {author} {\bibinfo {author} {\bibfnamefont {R.~M.}\ \bibnamefont
  {Diamond}}, \bibinfo {author} {\bibfnamefont {E.}~\bibnamefont {Matthias}},
  \bibinfo {author} {\bibfnamefont {J.~O.}\ \bibnamefont {Newton}}, \ and\
  \bibinfo {author} {\bibfnamefont {F.~S.}\ \bibnamefont {Stephens}},\ }\href
  {\doibase 10.1103/PhysRevLett.16.1205} {\bibfield  {journal} {\bibinfo
  {journal} {Phys. Rev. Lett.}\ }\textbf {\bibinfo {volume} {16}},\ \bibinfo
  {pages} {1205} (\bibinfo {year} {1966})}\BibitemShut {NoStop}%
\bibitem [{\citenamefont {Yamazaki}(1967)}]{YAMAZAKI19671}%
  \BibitemOpen
  \bibfield  {author} {\bibinfo {author} {\bibfnamefont {T.}~\bibnamefont
  {Yamazaki}},\ }\href {\doibase https://doi.org/10.1016/S0550-306X(67)80002-8}
  {\bibfield  {journal} {\bibinfo  {journal} {Nuclear Data Sheets. Section A}\
  }\textbf {\bibinfo {volume} {3}},\ \bibinfo {pages} {1} (\bibinfo {year}
  {1967})}\BibitemShut {NoStop}%
\bibitem [{\citenamefont {Rose}\ and\ \citenamefont
  {Brink}(1967)}]{Rose_Brink_Revmodphys_1967}%
  \BibitemOpen
  \bibfield  {author} {\bibinfo {author} {\bibfnamefont {H.~J.}\ \bibnamefont
  {Rose}}\ and\ \bibinfo {author} {\bibfnamefont {D.~M.}\ \bibnamefont
  {Brink}},\ }\href {\doibase 10.1103/RevModPhys.39.306} {\bibfield  {journal}
  {\bibinfo  {journal} {Rev. Mod. Phys.}\ }\textbf {\bibinfo {volume} {39}},\
  \bibinfo {pages} {306} (\bibinfo {year} {1967})}\BibitemShut {NoStop}%
\bibitem [{\citenamefont {Steffen}\ \emph {et~al.}(1975)\citenamefont
  {Steffen}, \citenamefont {Adler},\ and\ \citenamefont
  {Hamilton~(Ed.)}}]{Hamilton1975_electromagnetic}%
  \BibitemOpen
  \bibfield  {author} {\bibinfo {author} {\bibfnamefont {R.}~\bibnamefont
  {Steffen}}, \bibinfo {author} {\bibfnamefont {K.}~\bibnamefont {Adler}}, \
  and\ \bibinfo {author} {\bibfnamefont {W.}~\bibnamefont {Hamilton~(Ed.)}},\
  }\enquote {\bibinfo {title} {The electromagnetic interaction in nuclear
  spectroscopy},}\ \ (\bibinfo  {publisher} {North-Holland, Amsterdam},\
  \bibinfo {year} {1975})\BibitemShut {NoStop}%
\bibitem [{\citenamefont {Chalil}\ \emph {et~al.}(2022)\citenamefont {Chalil},
  \citenamefont {Materna}, \citenamefont {Litaize}, \citenamefont {Chebboubi},\
  and\ \citenamefont {Gunsing}}]{Chalil2022}%
  \BibitemOpen
  \bibfield  {author} {\bibinfo {author} {\bibfnamefont {A.}~\bibnamefont
  {Chalil}}, \bibinfo {author} {\bibfnamefont {T.}~\bibnamefont {Materna}},
  \bibinfo {author} {\bibfnamefont {O.}~\bibnamefont {Litaize}}, \bibinfo
  {author} {\bibfnamefont {A.}~\bibnamefont {Chebboubi}}, \ and\ \bibinfo
  {author} {\bibfnamefont {F.}~\bibnamefont {Gunsing}},\ }\href
  {https://doi.org/10.1140/epja/s10050-022-00683-0} {\bibfield  {journal}
  {\bibinfo  {journal} {The European Physical Journal A}\ }\textbf {\bibinfo
  {volume} {58}},\ \bibinfo {pages} {30} (\bibinfo {year} {2022})}\BibitemShut
  {NoStop}%
\bibitem [{\citenamefont {Piau}\ \emph {et~al.}(2023)\citenamefont {Piau},
  \citenamefont {Litaize}, \citenamefont {Chebboubi}, \citenamefont
  {Oberstedt}, \citenamefont {G\"o\"ok},\ and\ \citenamefont
  {Oberstedt}}]{PIAU2023137648}%
  \BibitemOpen
  \bibfield  {author} {\bibinfo {author} {\bibfnamefont {V.}~\bibnamefont
  {Piau}}, \bibinfo {author} {\bibfnamefont {O.}~\bibnamefont {Litaize}},
  \bibinfo {author} {\bibfnamefont {A.}~\bibnamefont {Chebboubi}}, \bibinfo
  {author} {\bibfnamefont {S.}~\bibnamefont {Oberstedt}}, \bibinfo {author}
  {\bibfnamefont {A.}~\bibnamefont {G\"o\"ok}}, \ and\ \bibinfo {author}
  {\bibfnamefont {A.}~\bibnamefont {Oberstedt}},\ }\href {\doibase
  https://doi.org/10.1016/j.physletb.2022.137648} {\bibfield  {journal}
  {\bibinfo  {journal} {Physics Letters B}\ }\textbf {\bibinfo {volume}
  {837}},\ \bibinfo {pages} {137648} (\bibinfo {year} {2023})}\BibitemShut
  {NoStop}%
\bibitem [{\citenamefont {Stuchbery}\ and\ \citenamefont
  {Robinson}(2002)}]{STUCHBERY2002753}%
  \BibitemOpen
  \bibfield  {author} {\bibinfo {author} {\bibfnamefont {A.~E.}\ \bibnamefont
  {Stuchbery}}\ and\ \bibinfo {author} {\bibfnamefont {M.~P.}\ \bibnamefont
  {Robinson}},\ }\href {\doibase https://doi.org/10.1016/S0168-9002(01)02114-3}
  {\bibfield  {journal} {\bibinfo  {journal} {Nuclear Instruments and Methods
  in Physics Research Section A: Accelerators, Spectrometers, Detectors and
  Associated Equipment}\ }\textbf {\bibinfo {volume} {485}},\ \bibinfo {pages}
  {753} (\bibinfo {year} {2002})}\BibitemShut {NoStop}%
\bibitem [{\citenamefont {Robinson}\ and\ \citenamefont
  {Stuchbery}(2002)}]{ROBINSON2002469}%
  \BibitemOpen
  \bibfield  {author} {\bibinfo {author} {\bibfnamefont {M.~P.}\ \bibnamefont
  {Robinson}}\ and\ \bibinfo {author} {\bibfnamefont {A.~E.}\ \bibnamefont
  {Stuchbery}},\ }\href {\doibase
  https://doi.org/10.1016/S0168-9002(02)00798-2} {\bibfield  {journal}
  {\bibinfo  {journal} {Nuclear Instruments and Methods in Physics Research
  Section A: Accelerators, Spectrometers, Detectors and Associated Equipment}\
  }\textbf {\bibinfo {volume} {489}},\ \bibinfo {pages} {469} (\bibinfo {year}
  {2002})}\BibitemShut {NoStop}%
\bibitem [{\citenamefont {Ferentz}\ and\ \citenamefont
  {Rosenzweig}(1955)}]{Ferentz_F_coeff}%
  \BibitemOpen
  \bibfield  {author} {\bibinfo {author} {\bibfnamefont {M.}~\bibnamefont
  {Ferentz}}\ and\ \bibinfo {author} {\bibfnamefont {N.}~\bibnamefont
  {Rosenzweig}},\ }\href@noop {} {\bibfield  {journal} {\bibinfo  {journal}
  {Available from Clearinghouse for Federal Scientific and Technical
  Information, U.S. Dept. of Commerce, Springfield, Va. 22151}\ } (\bibinfo
  {year} {1955})}\BibitemShut {NoStop}%
\bibitem [{\citenamefont {Hager}\ and\ \citenamefont
  {Seltzer}(1969)}]{HAGER19691_part3}%
  \BibitemOpen
  \bibfield  {author} {\bibinfo {author} {\bibfnamefont {R.}~\bibnamefont
  {Hager}}\ and\ \bibinfo {author} {\bibfnamefont {E.}~\bibnamefont
  {Seltzer}},\ }\href {\doibase https://doi.org/10.1016/S0550-306X(69)80002-9}
  {\bibfield  {journal} {\bibinfo  {journal} {Nuclear Data Sheets. Section A}\
  }\textbf {\bibinfo {volume} {6}},\ \bibinfo {pages} {1} (\bibinfo {year}
  {1969})}\BibitemShut {NoStop}%
\bibitem [{\citenamefont {Hager}\ and\ \citenamefont
  {Seltzer}(1968)}]{HAGER1968397}%
  \BibitemOpen
  \bibfield  {author} {\bibinfo {author} {\bibfnamefont {R.}~\bibnamefont
  {Hager}}\ and\ \bibinfo {author} {\bibfnamefont {E.}~\bibnamefont
  {Seltzer}},\ }\href {\doibase https://doi.org/10.1016/S0550-306X(68)80017-5}
  {\bibfield  {journal} {\bibinfo  {journal} {Nuclear Data Sheets. Section A}\
  }\textbf {\bibinfo {volume} {4}},\ \bibinfo {pages} {397} (\bibinfo {year}
  {1968})}\BibitemShut {NoStop}%
\bibitem [{\citenamefont {Raeside}\ \emph {et~al.}(1969)\citenamefont
  {Raeside}, \citenamefont {Ludington}, \citenamefont {Reidy},\ and\
  \citenamefont {Wiedenbeck}}]{RAESIDE1969677}%
  \BibitemOpen
  \bibfield  {author} {\bibinfo {author} {\bibfnamefont {D.}~\bibnamefont
  {Raeside}}, \bibinfo {author} {\bibfnamefont {M.}~\bibnamefont {Ludington}},
  \bibinfo {author} {\bibfnamefont {J.}~\bibnamefont {Reidy}}, \ and\ \bibinfo
  {author} {\bibfnamefont {M.}~\bibnamefont {Wiedenbeck}},\ }\href {\doibase
  https://doi.org/10.1016/0375-9474(69)90878-1} {\bibfield  {journal} {\bibinfo
   {journal} {Nuclear Physics A}\ }\textbf {\bibinfo {volume} {130}},\ \bibinfo
  {pages} {677} (\bibinfo {year} {1969})}\BibitemShut {NoStop}%
\bibitem [{\citenamefont {Robinson}(1990)}]{ROBINSON1990386}%
  \BibitemOpen
  \bibfield  {author} {\bibinfo {author} {\bibfnamefont {S.}~\bibnamefont
  {Robinson}},\ }\href {\doibase https://doi.org/10.1016/0168-9002(90)90395-M}
  {\bibfield  {journal} {\bibinfo  {journal} {Nuclear Instruments and Methods
  in Physics Research Section A: Accelerators, Spectrometers, Detectors and
  Associated Equipment}\ }\textbf {\bibinfo {volume} {292}},\ \bibinfo {pages}
  {386} (\bibinfo {year} {1990})}\BibitemShut {NoStop}%
\bibitem [{\citenamefont {Stetcu}\ \emph {et~al.}(2021)\citenamefont {Stetcu},
  \citenamefont {Lovell}, \citenamefont {Talou}, \citenamefont {Kawano},
  \citenamefont {Marin}, \citenamefont {Pozzi},\ and\ \citenamefont
  {Bulgac}}]{Stetcu_PhysRevLett.127.222502}%
  \BibitemOpen
  \bibfield  {author} {\bibinfo {author} {\bibfnamefont {I.}~\bibnamefont
  {Stetcu}}, \bibinfo {author} {\bibfnamefont {A.~E.}\ \bibnamefont {Lovell}},
  \bibinfo {author} {\bibfnamefont {P.}~\bibnamefont {Talou}}, \bibinfo
  {author} {\bibfnamefont {T.}~\bibnamefont {Kawano}}, \bibinfo {author}
  {\bibfnamefont {S.}~\bibnamefont {Marin}}, \bibinfo {author} {\bibfnamefont
  {S.~A.}\ \bibnamefont {Pozzi}}, \ and\ \bibinfo {author} {\bibfnamefont
  {A.}~\bibnamefont {Bulgac}},\ }\href
  {https://link.aps.org/doi/10.1103/PhysRevLett.127.222502} {\bibfield
  {journal} {\bibinfo  {journal} {Phys. Rev. Lett.}\ }\textbf {\bibinfo
  {volume} {127}},\ \bibinfo {pages} {222502} (\bibinfo {year}
  {2021})}\BibitemShut {NoStop}%
\end{thebibliography}%

\end{document}